\documentclass[fleqn,usenatbib]{mnras}
\usepackage{newtxtext,newtxmath}
\usepackage{graphicx} 
\usepackage{subcaption}
\usepackage{amsmath} 
\usepackage{xcolor}
\usepackage{float}
\usepackage{caption}
\usepackage{placeins}
\usepackage{lipsum}
\usepackage{multicol}
\usepackage{soul}
\usepackage{ragged2e}
\usepackage{threeparttable}

\def\refjnl#1{{\rm#1}}
\def\araa{\refjnl{ARA\&A}}             
\def\apj{\refjnl{ApJ}}                 
\def\apjl{\refjnl{ApJ}}                
\def\apjs{\refjnl{ApJS}}               
\def\ao{\refjnl{Appl.~Opt.}}           
\def\aap{\refjnl{A\&A}}                
\def\baas{\refjnl{BAAS}}               
\def\mnras{\refjnl{MNRAS}}             
\def\physrep{\refjnl{Phys.~Rep.}}   

\def\be{\begin{equation}} 
\def\ee{\end{equation}}

\def\kms{\,{\rm {km\, s^{-1}}}} 
\def\msun{{\Msun}}

\def\HI{\hbox{H~$\scriptstyle\rm I\ $}}

\def\gsim{\lower.5ex\hbox{\gtsima}} 
\def\lsim{\lower.5ex\hbox{\ltsima}} \def\gtsima{$\; \buildrel > \over 
\sim \;$} \def\ltsima{$\; \buildrel < \over \sim \;$} \def\prosima{$\; 
\buildrel \propto \over \sim \;$} \def\gsim{\lower.5ex\hbox{\gtsima}} 
\def\lsim{\lower.5ex\hbox{\ltsima}} 
\def\simgt{\lower.5ex\hbox{\gtsima}} 
\def\simlt{\lower.5ex\hbox{\ltsima}} 
\def\simpr{\lower.5ex\hbox{\prosima}} \def\la{\lsim} \def\ga{\gsim} 
  
\def\gtsima{$\; \buildrel > \over \sim \;$} 
\def\ltsima{$\; \buildrel < \over \sim \;$} 
\def\gsim{\lower.5ex\hbox{\gtsima}} 
\def\lsim{\lower.5ex\hbox{\ltsima}} 
\def\simgt{\lower.5ex\hbox{\gtsima}} 
\def\simlt{\lower.5ex\hbox{\ltsima}} 
\def\simpr{\lower.5ex\hbox{\prosima}} 
\def\la{\lsim} 
\def\ga{\gsim}

\def\msun{\,{\rm \Msun}}

\def\E3{{\cal E}_{\rm g}^{III}}

\def\msun{\rm M_\odot}

\def\M*{M_*}

\def\Z*{Z_*}
\def\L*{L_*}
\def\muv{\rm M_{UV}}

\def\rvir{R_{\rm vir}}
\def\rhil{R_{\rm hill}}
\def\mh{M_{\rm h}}
\def\mhacc{M_{\rm h}^{\rm acc}}

\def\msmin{M_*^{\rm min}}
\def\msmaj{M_*^{\rm maj}}
\def\cube{(2~{\rm cMpc})^3}
\def\agems{\rm Age_{M_{\star}}}

\def\mdark mattersa{M_{dark matter}^{sa}}

\def\env{{\rm log}(1+\delta)}
\def\mgi{M_\mathrm{g}^{\rm i}(z)}
\def\mgf{M_\mathrm{g}^{\rm f}(z)}

\title[Environmental-dependent galaxy assembly in the EoR]{Astraeus VII: The environmental-dependent assembly of galaxies in the Epoch of Reionization}
\author[Legrand et al.]{Laurent Legrand$^{1}$\thanks{e-mail: legrand@astro.rug.nl}, Pratika Dayal$^{1}$, Anne Hutter$^{1}$, Stefan Gottl\"ober$^2$, Gustavo Yepes$^{3,4}$ 
\newauthor and Maxime Trebitsch$^{1}$
\\
$^{1}$Kapteyn Astronomical Institute, University of Groningen, P.O Box 800, 9700 AV Groningen, The Netherlands\\
$^{2}$Leibniz-Institut f\"ur Astrophysik, An der Sternwarte 16, 14482 Potsdam, Germany\\
$^{3}$Departamento de Fısica Teorica, Modulo 8, Facultad de Ciencias, Universidad Autonoma de Madrid, 28049 Madrid, Spain\\
$^{4}$CIAFF, Facultad de Ciencias, Universidad Autonoma de Madrid, 28049 Madrid, Spain\\
}

\date{Accepted XXX. Received YYY; in original form ZZZ}

\pubyear{2022}

\begin{document} 
 
\date{} 

\maketitle

\begin{abstract}
Using the \textsc{astraeus} (semi-numerical rAdiative tranSfer coupling of galaxy formaTion and Reionization in N-body dark matter simUlationS) framework, we explore the impact of environmental density and radiative feedback on the assembly of galaxies and their host halos during the Epoch of Reionization. The \textsc{astraeus} framework allows us to study the evolution of galaxies with masses ($10^{8.2}\msun < \mh < 10^{13}\msun$) in wide variety of environment ($-0.5 < \env < 1.3$ averaged over $\cube$). We find that: (i) there exists a mass- and redshift- dependent "characteristic" environment (${\rm log} (1+\delta_a(\mh, z)) = 0.021\times (\mh/\msun)^{0.16} + 0.07 z -1.12$, up to $z\sim 10$) at which galaxies are most efficient at accreting dark matter, e.g at a rate of $0.2\%$ of their mass every Myr at $z=5$; (ii) the number of minor and major mergers and their contributions to the dark matter assembly increases with halo mass at all redshifts and is mostly independent of the environment; (iii) at $z=5$ minor mergers contribute slightly more (by up to $\sim 10\%$) to the dark matter assembly while for the stellar assembly, major mergers dominate the contribution from minor mergers for $\mh \lesssim 10^{11.5}\msun$ galaxies; (iv) radiative feedback quenches star formation more in low-mass galaxies ($\mh \lesssim 10^{9.5}\msun$) in over-dense environments ($\env > 0.5$); dominated by their major branch, this yields star formation histories biased towards older ages with a slower redshift evolution.
\end{abstract}



\section{Introduction} \label{Intro}
The Epoch of Reionization (EoR) marks the phase when the ionizing radiation emitted by the first galaxies drove the transition of the intergalactic medium (IGM) from a neutral and opaque to an ionized and transparent state \citep[see reviews by e.g.][]{barkana2001, dayal2018}. This phase transition had a significant effect on the formation of galaxies and the evolution of their observable properties at later times \citep[e.g.][]{weinmann2007, ocvirk2011, aubert2018, hutter2021a}. Indeed, the nature and evolution of galaxies were not only shaped by their dark matter mass assembly histories but also by the timing of reionization in their local region. Both processes have been shown to depend on the environmental density of a galaxy. Interpreting forthcoming observations of high-redshift galaxies will require an in-depth understanding of how different physical processes are tied to the evolution and properties of galaxies in regions of different environmental densities.

The underlying dark matter mass assembly histories of galaxies and their dependence on the density of their local environment have been investigated in various N-body simulations \citep[e.g.][]{gottloeber2001, gottloeber2002, fakhouri2008, fakhouri2009, hahn2009, fakhouri2010}. In line with hierarchical structure formation, low-mass perturbations are found to collapse first from primordial Gaussian density fluctuations. The resultant bound dark matter structures form increasingly more massive halos as they both accrete matter from the IGM and merge over time. These two processes show competing trends with the environment.

On the one hand, the number of mergers of halos per unit of time effectively rises with the depth of the underlying gravitational potential. This translates into a rise of the merger rate with an increasing mass of the merged halo, increasing difference in progenitor halo masses, and increasing redshift \citep[e.g.][]{gottloeber2001, fakhouri2008, fakhouri2009, genel2010, rodriguezgomez2015, duncan2019, oleary2021}. Importantly, the higher abundance of halos in denser regions increases the probability of halos merging - this causes the merger rate and the merged mass to correlate positively with the local density \citep{fakhouri2010}. These insights have been gained from simulations since observationally, works are restricted to $z<3$ \citep[see e.g][]{bertone2009, bluck2012, mundy2015, ferreira2020, conselice2022} or very limited fields \citep[][]{ventou2017}.

On the other hand, the halo mass growth through smooth accretion depends on the local density in a more complex manner. A halo in the vicinity of a more massive halo shows lower accretion rates than a halo of the same mass that has no massive neighbors, i.e the mass growth through smooth accretion correlates negatively with an increase in the local density. This leads to halos in dense regions forming earlier than their equally massive counterparts in less dense regions, known as the assembly bias \citep[e.g.][]{gottloeber2002, sheth2004, harker2006, wechsler2006, gao2007, jing2007}. 
By analyzing dark matter-only (N-body) simulations \citet{hahn2009} explained this effect as follows: the tidal forces generated by the gravitational potential of neighboring massive halos induce a velocity shear around lower mass halos. This changes the convergence of the accretion flows onto the lower-mass halos and suppresses their accretion. As gravity enhances the density of over-dense regions over time, this velocity shear grows. Together with the hierarchical nature of structure formation, this also implies that a halo of a given mass has a characteristic density of the local environment in which its accretion rate is maximum. In more dense regions the given halo is subject to the tidal field of a more massive neighbour, while its probability of forming in less dense regions drops significantly.

\citet{fakhouri2010} investigated for which halos the mass growth is dominated by smooth accretion (mergers) and correlates negatively (positively) with the local density. Their analysis revealed that halos in denser regions show a higher mass growth rate via mergers and a lower mass growth rate via smooth accretion than equally massive halos in less dense regions. However, a caveat is that their analysis of the Millenium simulation focuses on lower redshifts ($z\lesssim2$) and more massive halos ($\mh\gtrsim5\times10^{10}\msun$) than are abundant during the EoR. It still remains an open question as to whether the massive galaxies during the EoR are similarly dominated by mass growth via mergers or assemble most of their mass through smooth accretion due to the overall large-scale mass distribution being more homogeneous.

The local-density-dependent halo mass assembly inherently shapes the evolution of galaxies by defining the underlying gravitational field that channels gas motions. But the extent to which the physical processes driving the evolution of galaxies propagate the dependence of the halo mass assembly on the local density into the evolution of their galactic properties, such as stellar masses and star formation histories (SFHs), remains an open question. If gas accretion is the dominant process of gaining mass, we would expect galaxies to exhibit lower star formation rates and shallower SFHs in denser regions given their gas accretion rates would be lower as compared to equally massive galaxies in less dense regions.
Feedback processes that correlate with star formation and regulate it, such as supernovae (SN) explosions, are unlikely to change this local density dependence, given their efficiency in reducing cold gas and star formation depends primarily on the depth of the gravitational potential. However, radiative (photoheating) feedback from the ultraviolet background (UVB) that is built up in ionized regions during reionization leads to the star formation rate in low-mass galaxies ($\mh\lesssim10^9\msun$) being dependent on the redshift when their local regions were ionized \citep{gnedin2000, hoeft2006, dawoodbhoy2018, hutter2021a}. The relation between the redshift when a region was ionized and its density depends on whether the ionization fronts propagate from over-dense to under-dense regions (inside-out reionization topology) or vice-versa (outside-in). Different works hint at the outside-in component dominating towards the end of the EoR \citep[][]{miralda2000, wyithe2003, choudhury2005}. However, most reionization simulations suggest that reionization proceeds overall in an inside-out fashion \citep[e.g.][]{iliev2006, iliev2012, choudhury2006, trac2007, battaglia2013, bauer2015, trebitsch2021}. Consequently, in the latter scenario, low-mass galaxies in dense regions will show a stronger suppression of their star formation rates than galaxies in less dense regions.

The negative correlation between star formation and local density due to gas accretion and radiative feedback from reionization could be diluted by mergers: as two galaxies with different masses but similar environmental densities merge, the lower mass galaxy will have a lower gas-to-dark matter ratio than the higher mass galaxy when compared to their equally massive counterparts in the regions of characteristic density. Thus, the resultant merged galaxy grows in dark matter mass but has a gas-to-dark matter ratio lower than its higher mass progenitor, which - if gas accretion does not fully compensate this gas loss - reduces the star formation compared to a scenario where gas accretion clearly dominates. 

To date, most galaxy observations that measure the trends of galactic properties with local density have been conducted at low redshifts ($z<2$). These hint at: {\it (i)} the average metallicity of star-forming galaxies with $M_\star\simeq10^{9.5-11}\msun$ being higher in over-dense regions than their equally massive field counterparts \citep{chartab2021}; and {\it (ii)} the gas-phase metallicity of satellite star-forming galaxies exceeding that of equally massive centrals \citep[e.g.][]{pasquali2010, pasquali2012, peng_maiolino2014, wu2017, lian2019, schaefer2019}; this is in agreement with hydrodynamical simulations that find a suppression of cold gas accretion in dense environments and satellites \citep{vandeVoort2017}. At intermediate redshifts ($z\simeq2-4$) the trend of the gas-phase metallicity of star-forming galaxies with local density seems to be inverted \citep{chartab2021, calabro2022} and star formation enhanced in dense regions \citep{lemaux2020}, suggesting e.g. more pristine gas inflows or outflows induced by an enhanced merger rate, or more efficient gas stripping and harassment.

Due to the limitation of high-redshift ($z\gtrsim6$) galaxy observations to the brightest objects, the environmental dependence of galactic properties at such redshifts has been mostly studied through simulations. However, most studies have focused on investigating the dependence of properties, such as the star formation rate, on the redshift when a region was reionized \citep{ocvirk2016, dawoodbhoy2018, hutter2021a, ocvirk2020}. By analyzing zoom-in hydrodynamical simulations of a large simulation box, \citet{lovell2021} found no dependence of the star-forming sequence on the density of the environment between z=5 and z=10, a result that has been found in other simulations, albeit at lower redshifts \citet{hirschmann2016, bassini2020, yajima2022}. A key challenge in investigating the environmental dependence of accretion and mergers during the EoR is that a simulation is needed that has: {\it (i)} a large enough volume to sample different environments; {\it (ii)} tracks the representative galaxy population; and {\it (iii)} accounts for radiative feedback from reionization. 

Our \textsc{astraeus} framework, which couples a state-of-the-art N-body simulation with a semi-analytical galaxy evolution model and a semi-numerical reionization scheme self-consistently, fulfills these criteria perfectly. In this work, we use {\sc astraeus} to assess how the assembly of galaxies depends on the density of their environment. In particular, we focus on the following questions: How does the environment (local density) affect the assembly of dark matter halos and the associated stellar mass? What is the relative role of major and minor progenitors in building both the halo and stellar contents? What is the impact of radiative feedback as a function of the halo mass and environmental overdensity?

This paper is structured as follows. In Sec.~\ref{Sec_Model}, we describe the {\sc astraeus} framework and outline the different radiative feedback models that describe the interplay between galaxy evolution and reionization. In Sec.~\ref{Sec_assembly}, we discuss the mass assembly of low-mass and massive galaxies in different environments. In Sec.~\ref{Sec_dark matter}, we investigate the role of the local over-density in the importance of the major branch in halo assembly. We then extend this analysis to the environment-dependent assembly of the stellar mass (Sec.~\ref{sec_star}). Finally, we conclude in Sec.~\ref{Sec_conclusion}.

\section{The model} \label{Sec_Model}
This paper is the seventh in a series of works that use the \textsc{astraeus} (semi-numerical rAdiation tranSfer coupling of galaxy formaTion and Reionization in N-body dArk mattEr simUlationS) framework. We briefly describe the model here and readers are referred to \cite{hutter2021a} for complete details. This framework couples the dark matter merger trees from the \textit{Very Small MultiDark Planck} (\textsc{vsmdpl}) N-body simulation\footnote{This has been run as part of the MultiDark project: www.cosmosim.org.} with a modified version of the {\textsc {delphi}} semi-analytic model for galaxy formation \citep{dayal2014} and the {\textsc {cifog}} semi-numerical scheme \citep{hutter2018} for reionization. The N-body simulation is based on a version of \textsc{gadget-2} Tree + PM \citep{springel2005} and is run for a box size of $160\,h^{-1}$Mpc using $3840^3$ particles. This results in a dark matter mass resolution mass of $6.2\times10^6\, h^{-1}\rm M_\odot$. The phase-space halo finder \textsc{rockstar} \citet{behroozi2012a} was used to extract halos and the merger trees were then derived using the \textsc{consistentree} algorithm \citet{behroozi2012b}. The halos used for this study have a minimum bound mass of $1.2\times 10^8\,h^{-1}\mathrm{M_\odot}$ using a minimum of 20 particles.

We use the first 74 simulation snapshots between $z=25$ and $z=4.5$ for our calculations. At every redshift snapshot, \textsc{Astraeus} tracks the evolution of galaxies as follows:

\begin{itemize}
    \item \textit{Dark Matter}: the dark matter mass of a halo at redshift $z$, $\mh(z)$, can be written as 
    \begin{equation}
        M_\mathrm{h}(z) = M_\mathrm{h}^\mathrm{prog}(z + \Delta z) + M_\mathrm{h}^\mathrm{acc}(z, z+\Delta z)
    \end{equation} 
    where $M_\mathrm{h}^\mathrm{prog}(z+\Delta z)$ is the halo mass brought in by merging progenitors at $z+\Delta z$ (equal to 0 if the galaxy has no progenitors) and $M_\mathrm{h}^\mathrm{acc}(z, \Delta z)$ is the accreted dark matter mass between $z+\Delta z$ and $z$. The major branch is identified as follows: for any given galaxy, we start by selecting the most massive progenitor, with mass $M_\mathrm{h}^\mathrm{maj}(z+\Delta z)$, at the previous redshift-step $z+\Delta z$. We then select the most massive progenitor of this galaxy at the previous redshift step. We repeat this process until we reach the beginning of the simulation or there are no more progenitors. All of these progenitors are considered to be part of the major branch and are called ``major progenitors'' while all other progenitors are ``minor progenitors''; the sum of the masses of the minor progenitors at any $z$ is expressed as $M^{\rm min}_{\rm h}(z)$. We note that this procedure is different from simply selecting the most massive progenitors of a galaxy at each redshift since it is possible for the major branch to start assembling later but faster than other branches. For an example, we refer the reader to Fig. 1 in \citet{legrand2021}. We note that halos with masses below our resolution limit will by construction not be considered as ``minor progenitors'': while this is completely negligible for high mass halos (for halos with $\mh \gtrsim 10^{10.2}\,\mathrm{M}_{\odot}$, these are mergers with 1:100 mass ratios), this is not entirely the case at low-mass, where this leads to an over-estimation of the smooth accretion component.

    \item \textit{Initial gas mass}: As for the dark matter mass, galaxies assemble their initial gas mass, $\mgi$, from both mergers and smooth accretion from the IGM such that
     \begin{equation}
        \mgi = M_\mathrm{g}^\mathrm{f,maj}(z+\Delta z) + M_\mathrm{g}^\mathrm{f,min}(z+\Delta z) + 
        \frac{\Omega_b}{\Omega_m} M_\mathrm{h}^\mathrm{acc}(z+\Delta z).
    \end{equation} 
   Here, the first two terms on the RHS show the final gas masses brought in by merging major and minor progenitors (if any), respectively, after star formation and Type II supernova (SNII) feedback. The third term shows the gas mass accreted from the IGM assuming that the accretion of dark matter drags in a cosmological baryon-to-dark matter ratio of gas mass. This gas mass can be reduced due to the impact of radiative feedback from reionization as detailed in what follows.
    
     \item \textit{Star formation and supernova feedback}:
    At each redshift z, we assume that a galaxy turns a fraction of its initial gas mass into stars with an \textit{effective} efficiency $f\mathrm{_\star^{eff}}$ expressed as 
    \begin{equation} \label{eq:feff}
        f_\mathrm{eff} = \mathrm{min}[f^\mathrm{ej}_\star, f_\star],
    \end{equation}
    where $f^\mathrm{ej}_\star$ is the fraction of gas that the galaxy can turn into stars before the SNII unbind the rest of the gas and $f_\star$ is a maximum threshold efficiency. Our SNII feedback model accounts for stellar lifetimes as a result of which $f^\mathrm{ej}_\star$ is expressed as
    \begin{equation}
        f^\mathrm{ej}_\star(z) = \frac{\upsilon_c^2}{\upsilon_c^2 + f_{\rm w} E_{51}\nu_z}\left[ 1 - \frac{f_{\rm w} E_{51}\sum_j \nu_j M^\mathrm{new}_{\star,j}(z_j)}{M^i_g(z)\upsilon_c^2}\right],
        \label{eq:fej}
    \end{equation}
    where $\upsilon_c$ is the rotational velocity of the halo, $E_{51} = 10^{51}\mathrm{erg\,s^{-1}}$ the energy produced by each SNII, $f_\mathrm{w}$ is the fraction of SNII energy that couples to the gas and drives the winds, $M^\mathrm{new}_{\star,j}(z_j)$ the newly-formed stellar mass in time step $z_\mathrm{j}$, and $\nu_j$ the corresponding fraction that explodes as SNII in time step $z$ using  a Salpeter IMF between $0.1 \mathrm{M_\odot}$ and $100 \mathrm{M_\odot}$. The gas mass left in a galaxy after star formation and SNII feedback is termed the final gas mass, $\mgf$. We do not include any ``burst''-mode enhancement of star formation due to mergers in the \textsc{Astraeus} framework: while mergers are expected to drive gas to the central regions of the galaxy and e.g. enhance merger activity \citep[see][for example]{croton2006}, previous semi-analytical work has suggested that this is not going to be a dominant contributor to the global star formation history \citep[e.g.][]{somerville2008,somerville2015}.

    \item \textit{Stellar mass assembly}:
    The amount of stars newly formed by a galaxy at redshift z is
    \begin{equation}
        M_\star^\mathrm{new}(z) = f_\mathrm{eff} * M_\mathrm{g}^\mathrm{i}(z)
        \label{eq:mstarnew}
    \end{equation}
   The total stellar mass at any redshift then is
    \begin{equation}
  M_\star(z) = M_\star^\mathrm{new}(z)+M_\star^{\rm maj}(z+\Delta z) + M_\star^{\rm min}(z+\Delta z),
  \end{equation}
  where the last two terms on the RHS show the stellar mass brought in by merging major and minor progenitors, respectively.
    
    \item \textit{Radiative feedback}:
   The \HI ionizing photons produced by star formation are obtained from \textsc{starburst99} \citep{leitherer1999} using the entire star formation history of a galaxy. A fraction of these can escape into the IGM ($f_{\rm esc}$) driving the process of reionization. If the cumulative number of ionizing photons emitted exceeds the cumulative number of absorption events, a region is considered ionized which is accompanied by an increase in the temperature. This can cause gas to photo-evaporate \citep{barkana1999, shapiro2004, iliev2005b} and the pressure to increase, resulting in a higher Jeans mass scale which leads to a reduction of gas infall \citep{gnedin2000, hoeft2006}. Thus, the initial gas mass available for star formation in a galaxy located in an ionized region can be expressed as 
    \begin{equation}
        M_\mathrm{g}^\mathrm{i}(z) = \mathrm{min}\left[ M_\mathrm{g}^\mathrm{acc}(z) + M_\mathrm{g}^\mathrm{prog}(z), f_\mathrm{g} \frac{\Omega_b}{\Omega_m}\mh\right],
    \end{equation}
    where $f_\mathrm{g}$ is the fraction of gas mass not affected by radiative feedback. In this work, we explore two radiative feedback models, whose strengths are defined by a characteristic mass at which $f_\mathrm{g}=0.5$, $M_c(z)$\footnote{The evolution of $M_c(z)$ in both model can be found in Fig. 1 of \cite{hutter2021a}.}. The first is the \textit{Photoionization} model where the strength of the feedback increases with an increase in the photoionization rate and the difference between the reionization redshift and the current galaxy redshift. The second (maximal feedback) model, called the \textit{Jeans Mass} model, assumes the gas density to react instantaneously to the gas temperature increasing to $T_0=4\times 10^4\,\rm K$ in ionized regions. 
    
\end{itemize}
Our model has two mass- and redshift-independent free parameters ($f_\star$ and $f_\mathrm{w}$) that are tuned by simultaneously matching to the evolving ultraviolet luminosity function and stellar mass function at $z \sim 5-10$. The third free parameter ($f_{\rm esc}$) is tuned so as to reproduce the key reionization observables such as the electron scattering optical depth and the constraints on the ionization state inferred from Lyman Alpha emitters, Gamma-ray bursts and quasars\footnote{More details on the tuning of these parameters can be found in Section 3 from \citep{hutter2021a}.}.

Throughout the paper, we use the following cosmological parameters: [$\rm \Omega_\Lambda$, $\rm \Omega_m$, $\rm \Omega_b$, $\rm h$, $\rm n_s$, $\rm \sigma_8$] = [0.69, 0.31, 0.048, 0.68, 0.96, 0.82]. Finally, in what follows, at any redshift the environmental over-density is calculated as $1+\delta(z) = \rho(z) / {\Bar \rho(z)}$, where ${\Bar \rho(z)}$ is the mean density of the simulation box at redshift $z$ and $\rho(z)$ is the density averaged over a cube of $\cube$. We choose a smoothing scale of $\cube$ as it is the mean distance between galaxies with $\muv \simeq -16$ at $z\simeq7$, corresponding to the limiting magnitude using the NIRCam imaging in the JADES GTO survey quoted by \cite{williams2018} and \cite{rieke2019}.

\section{The environmental dependence of the mass assembly of early galaxies} \label{Sec_assembly}
We start by looking at the assembly of the dark matter, gas and stellar mass components of $z \sim 5$ galaxies with halo masses ranging between $\mh \sim 10^{9.5-11.5}\msun$ in under- and over-dense environments. Our aim is to answer two key questions: {\it (i):} what is the role of mergers versus accretion in their assembly; and {\it (ii):} how does the assembly depend on the environmental density? In what follows, the environmental density bins have been chosen to ensure that we sample the entire density range occupied by galaxies of the chosen halo mass whilst having a sufficient number of galaxies to obtain statistically significant results.

\begin{figure*}
    \centering
    \includegraphics[width=\textwidth]{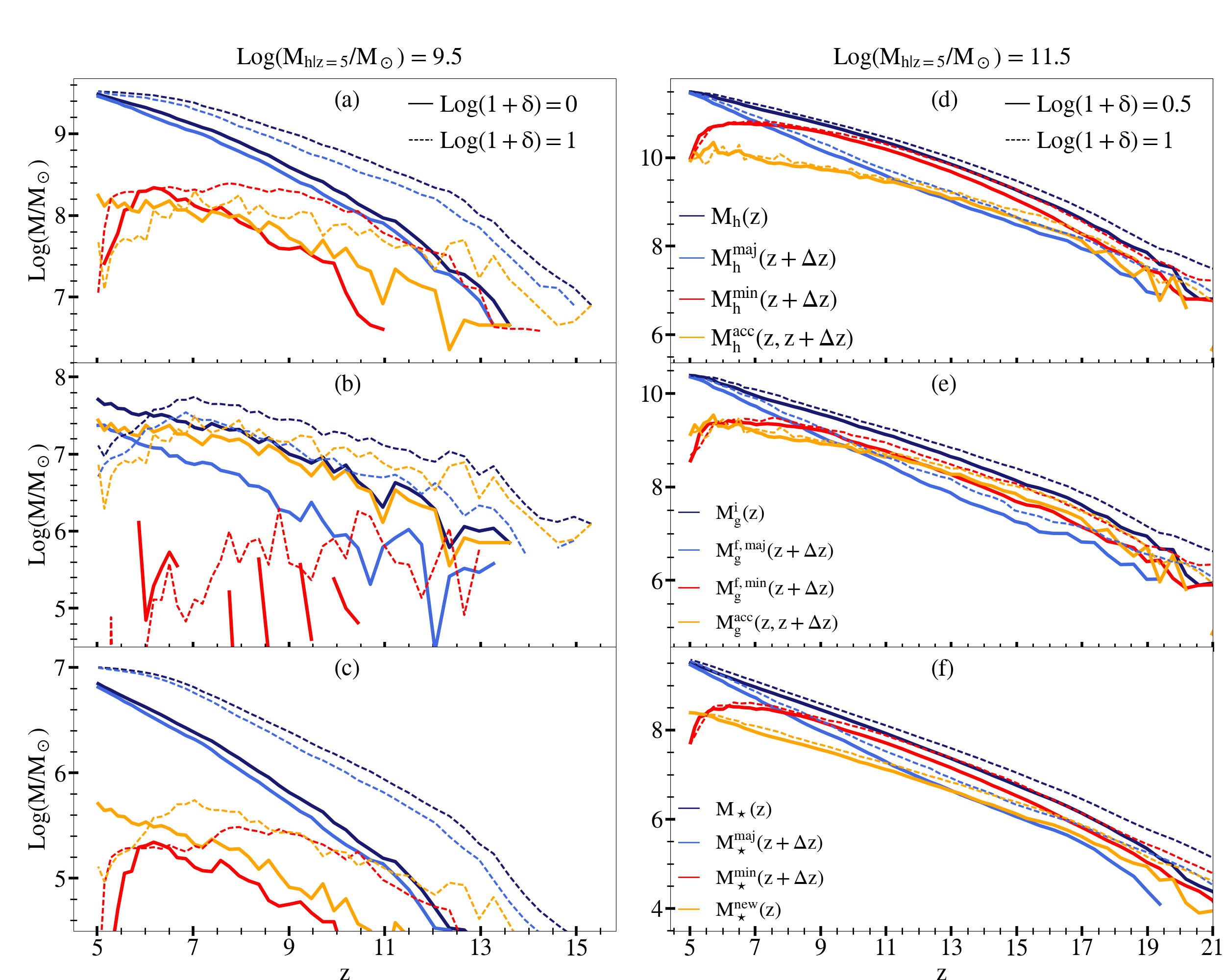}
    \caption{Galaxy assembly as a function of redshift and environmental over-density in the \textit{Photoionization} model. Each column shows the assembly averaged over 80 galaxies with the $z=5$ halo mass indicated at the top: these include 40 galaxies each in low- and high-density environments (as marked) shown using the solid and dotted lines, respectively. The rows show the assembly of: the halo mass (\textit{first row}), gas mass (\textit{middle row}) and stellar mass (\textit{bottom row}). As marked, for each component, we show the masses summed over all progenitors at redshift $z$ (black line), the mass in the major and minor progenitors at the previous redshift step $z+\Delta z$ (blue and red lines, respectively) and the smoothly-accreted dark matter and gas masses and the newly formed stellar mass between $z+\Delta z$ and $z$ (yellow line).} 
    \label{fig:Fig1}
\end{figure*}

We start by looking at the assembly of low-mass galaxies ($\mh \sim 10^{9.5}\msun$) in panel (a) from Fig. \ref{fig:Fig1}. As seen, in regions of critical density where $\rho(z) = {\Bar \rho(z)}$, such galaxies start assembling by $z \sim 10$. Their assembly is quasi-monolithic, being dominated by the major branch progenitor that contains $\gsim 80\%$ of the dark matter mass at almost all redshifts. Minor progenitors contribute less than 10\% to the mass build-up at any redshift, with their importance decreasing with decreasing redshift, as these merge into the final $z \sim 5$ halo. We also find smooth accretion from the IGM to be as important as mergers (at the 10\% level) in determining the assembly of such low-mass halos in average-density environments. We note however that this is an upper-limit to the contribution of smooth accretion, since in this low-mass regime, the contribution of mergers just below the resolution limit is counted as smooth accretion.

Low-mass halos in over-dense environments, with $\rho(z) = 13.5 {\Bar \rho(z)}$, start assembling earlier (at $z \sim 15.5$) and reach the same mass as their counterparts in low-density regions earlier by $\Delta z \sim 2$, throughout their assembly history. The assembly, however, is qualitatively similar to halos of similar masses in low-density environments. Here too, the major branch dominates the mass assembly, bringing in $\gsim 70\%$ of the mass with minor mergers and IGM smooth-accretion bringing in $\sim 15\%$ of the mass at any redshift. Interestingly, however, we find that while smooth accretion is higher at high-redshifts ($z \gsim 6.5$) in high-density regions, at lower redshifts this trend reverses: smooth accretion in low-density regions overtakes that in higher-density regions. This is driven by the competing tidal force from a larger number of neighboring galaxies in high-density environments, as discussed in Sec. \ref{global_dark matter} that follows. 

We then discuss the assembly of the gas mass associated with such low-mass halos as shown in panel (b) of the same figure. As discussed in Sec. \ref{Sec_Model}, the initial gas mass at the beginning of a redshift step is determined both by the final gas mass brought in by merging progenitors (after star formation and SNII feedback) and the smoothly-accreted gas from the IGM; both these components are affected by reionization feedback if the galaxy lies in an ionized environment as also detailed in Sec. \ref{Sec_Model}. The progenitors of such low-mass halos in average density regions mostly form stars in the SNII-feedback limited regime (i.e. at a given redshift, SNII can push out almost all of the gas mass out of the halo potential, quenching further star formation) as a result of which most of the gas mass ($\gsim 80\%$) is assembled through smooth accretion of gas from the IGM. As the major branch builds up its mass, its contribution to the gas content rises from $\lsim 10\%$ at $z \gsim 7$ to $\sim 40\%$ at lower redshifts. As might be expected, given their low masses, minor progenitors only have a negligible contribution to assembling the gas mass. 

As seen from panel (b), in high-density regions too, smooth accretion dominates the gas mass assembly. However, tracking the accreted dark matter mass, the smoothly-accreted IGM gas also shows a decrease at $z \lsim 6.5$. Further, the initial gas mass for such galaxies shows a decrease at $z \lsim 6$ - this is possibly driven by the impact of radiative feedback. Indeed, as shown in Fig. \ref{fig:zreion}, while the environments of such low-mass halos in average density regions only get reionized as late as $z \sim 8-10$, they are reionized much earlier (by $z \sim 14-16$) in high-density environments. As a result, radiative feedback is more effective in preferentially reducing the gas mass of low-mass galaxies in over-dense regions. As a result of its mass build-up and the competing tidal pull on gas from neighboring galaxies, the major branch over-takes the gas mass contributed by smooth accretion at $z \lsim 9$; minor progenitors contribute $\lsim 10\%$ to the gas mass at any redshift.

We then look at the assembly of the stellar mass in panel (c) of the same figure. In average-density environments, most ($\gsim 80\%$) of the stellar mass for low-mass galaxies is brought in by the major branch with (SNII-feedback limited) minor progenitors bringing in less than $10\%$ of the stellar mass. The newly formed stellar mass increases with decreasing redshift and is mostly driven by star formation in the major branch which increases as it assembles its halo potential (and the associated gas mass). Galaxies in high-density environments show similar qualitative trends: a key difference is that while minor progenitors contribute a significant amount ($\sim 40\%$) to the stellar mass at $z \gsim 12$, the major progenitor rapidly takes over at lower redshifts as minor progenitors rapidly merge into the final halo assembling at $z \sim 5$. 

We then discuss the assembly of high-mass $z \sim 5$ galaxies (with $\mh \sim 10^{11.5}\msun$) as shown in the right column of Fig. \ref{fig:Fig1}. Firstly, given that these correspond to $> 5-\sigma$ fluctuations, such galaxies are preferentially found in high-density regions. As shown, we explore their assembly in regions that are moderately and highly over-dense with $\rho(z) = 4 {\Bar \rho(z)}$ and $10 {\Bar \rho(z)}$, respectively. We find the assembly of such high-mass galaxies to be much more complex, being shaped by a combination of the mass being brought in by the major branch, minor progenitors and smooth accretion. Additionally, contrary to the low-mass galaxies discussed above, most of the mass is brought in by a large number of low-mass (minor) progenitors for most of their assembly history as now discussed and shown in panel (d) of this figure. In moderately over-dense regions, such galaxies start assembling at $z \sim 21$. Down to $z \sim 17.5$, the halo mass assembly is equally driven by minor progenitors and smooth accretion. As an increasing number of low-mass progenitors form and merge into the assembling potential, their contribution to the mass increases from $\sim 50\%$ at $z \sim 17.5$ to $\sim 80\%$ by $z \sim 11.8$. Between $z \sim 17.5-11.8$, the major branch and smooth-accretion both contribute roughly equally ($\sim 10\%$) to the mass assembly. Below $z \sim 11.8$, the assembly of the major branch accelerates as it builds up its potential and it becomes as important as the sum of the mass locked up in the minor branches ($\sim 45\%$ of the total mass) by $z \sim 6.8$. Below this redshift, the major branch dominates the assembly down to $z \sim 5$. The situation is qualitatively similar for high-mass galaxies in highly over-dense regions where minor progenitors again dominate the halo mass assembly between the formation redshift of the first progenitors ($z \sim 23$) down to $z \sim 7.5$. In the initial phases ($z \sim 23-13.7$), minor progenitors become increasingly important while the major branch and smooth-accretion are roughly equally important. By $z \sim 13.7$, minor progenitors bring in $\sim 70\%$ of the mass with the major branch and accretion contributing at the order of $15\%$. As the mass in the major progenitor grows, it starts being as important as the other minor progenitors such that both contribute roughly $\sim 42\%$ to the halo mass by $z \sim 7.5$; below this redshift, the assembly is dominated by the major branch.

\begin{figure*}
    \centering
    \includegraphics[width=\textwidth]{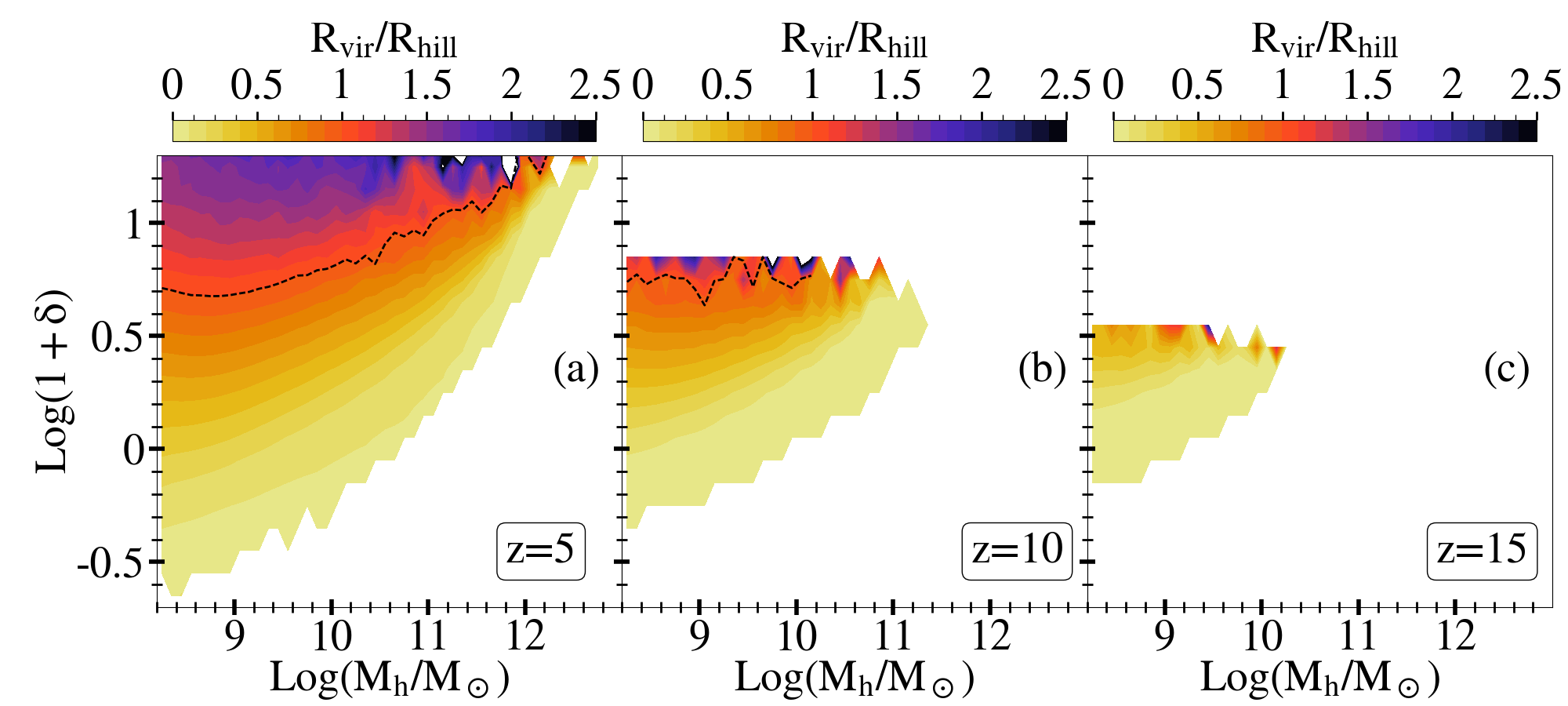}
    \caption{We show the median strongest tidal force from nearby halos, expressed using the dimensionless parameter ($\rvir/\rhil$), as a function of halo mass and density of the environment averaged over $\cube$ at $z=15, 10, 5$ (as marked) for the \textit{Photoionization} model. The colorbar shows the ratio $\rvir/\rhil$ and the ratio of 1 is shown by the dotted line.} 
    \label{fig:tidal}
\end{figure*}

We then discuss the build-up of the gas mass for such high-mass halos. In moderately over-dense regions, minor mergers and smooth accretion are equally important and contribute roughly $50\%$ to building the gas mass at $z \gsim 17.5$. Although accretion is sub-dominant to minor mergers in building up the gas mass between $z \sim 17.5-11.8$, as a result of gas suppression due to feedback (both SN and reionization), both keep contributing equally to the gas mass down to $z \sim 11.8$. As the major branch assembly accelerates, it overtakes gas brought in by smooth accretion at $z \sim 11.8$ - at this point, major mergers bring in $\sim 65\%$ of the gas mass, with minor mergers and accretion bringing in $\sim 18\%$. As the major progenitor establishes as the dominant mass component, it overtakes the contribution from minor progenitors at $z \lsim 8$. Below this redshift, the major progenitor contributes $\gsim 75\%$ of the gas mass, followed by minor progenitors ($\lsim 15\%$) and accretion ($\lsim 10\%$). The situation is again qualitatively similar in high-density regions, with minor mergers and accretion contributing equally ($\sim 40\%$) to the gas mass down to $z \sim 12.5$; major progenitors contribute $\lsim 10\%$ to the gas mass by this point. As its mass assembles, the major branch overtakes the gas mass from accretion and minor progenitors at slightly higher redshifts of $z \sim 10$ and 8, respectively. 

We look at the assembly of the stellar mass for high-mass halos in panel (f) of the same figure. We find that, unlike low-mass halos where the major branch dominates the stellar mass assembly, for high-mass halos minor progenitors dominate the assembly of the stellar mass over most of the formation history between $z \sim 21-8$. While most of this is new star formation in the minor progenitors at the earliest epochs, the faster growth rate of the major branch, driven by a combination of accretion and mergers of minor progenitors (at $z \lsim 11.8$) results in its domination of the stellar mass assembly at $z \lsim 8$. Qualitatively, the assembly in high-density regions is very similar with minor progenitors driving the assembly at $z \gsim 8.75$ at which point the major branch starts dominating in terms of stellar mass. 


\section{The assembly of the halo mass and the role of the major branch} \label{Sec_dark matter}

In this section, we discuss the role of neighbouring halos on dark matter accretion in Sec.~\ref{global_dark matter}, dark matter accretion rates in Sec. \ref{acc_rate} and the accreted mass weighted ages in Sec. \ref{ages_halos} before ending with the importance of the major branch in Sec.~\ref{major}.

\subsection{Impact of neighboring halos on dark matter accretion} \label{global_dark matter}
In this section, we start by discussing the influence of neighboring galaxies on the dark matter accretion rate of a halo. To study this, we calculate the Hill radius which represents the contribution of a halo to the local tidal field with respect to more massive neighbors \citep[see e.g][]{hahn2009, rodriguezpuebla2017}. Since most halos have more than one neighbor, we use the minimum Hill radius which is an upper bound on the spatial extent of newly infalling material that can remain gravitationally bound to a halo. We then calculate the Hill radius for a given halo of mass $m$ with all halos such that
\begin{equation}
    \rhil=D\bigg(\frac{m}{3M}\bigg)^{1/3}. \label{eq:Hill}
\end{equation}
where $M$ and $D$ are the masses and distances to neighbouring halos. The minimum Hill radius is then the minimum value of the Hill radius obtained above. Although this Hill radius only accounts for the tidal field created by the most influential neighbor, our assumption is supported by the findings of e.g. \citet{hahn2009} who have shown that considering the tidally dominant neighbor alone is sufficient to accurately assess the effect of tides on the accretion on a given halo. To assess the capacity of a halo to accrete dark matter from the IGM, we compare its Hill radius to its virial radius $\rvir$\footnote{We define $\rvir$ as the radius within which the density is $\sim 178$ times the mean matter density.}.

In Fig.~\ref{fig:tidal}, we show the ratio of the Hill radius to the virial radius as a function of the halo mass and environmental density at $z=5-15$. While a value of $R_{\rm vir}/\rhil\ll 1$ implies that accretion onto a given halo is not affected by surrounding halos (either because it has no close neighbors or because the neighbors have low masses) a value of $R_{\rm vir}/\rhil\gsim 1$
(dotted line in Fig.~\ref{fig:tidal}) implies that in-falling material can only remain gravitationally bound to the halo if it is already located within it i.e. accretion on the halo can be completely halted by more massive neighboring halos.

\begin{figure*}
    \centering
    \includegraphics[width=\textwidth]{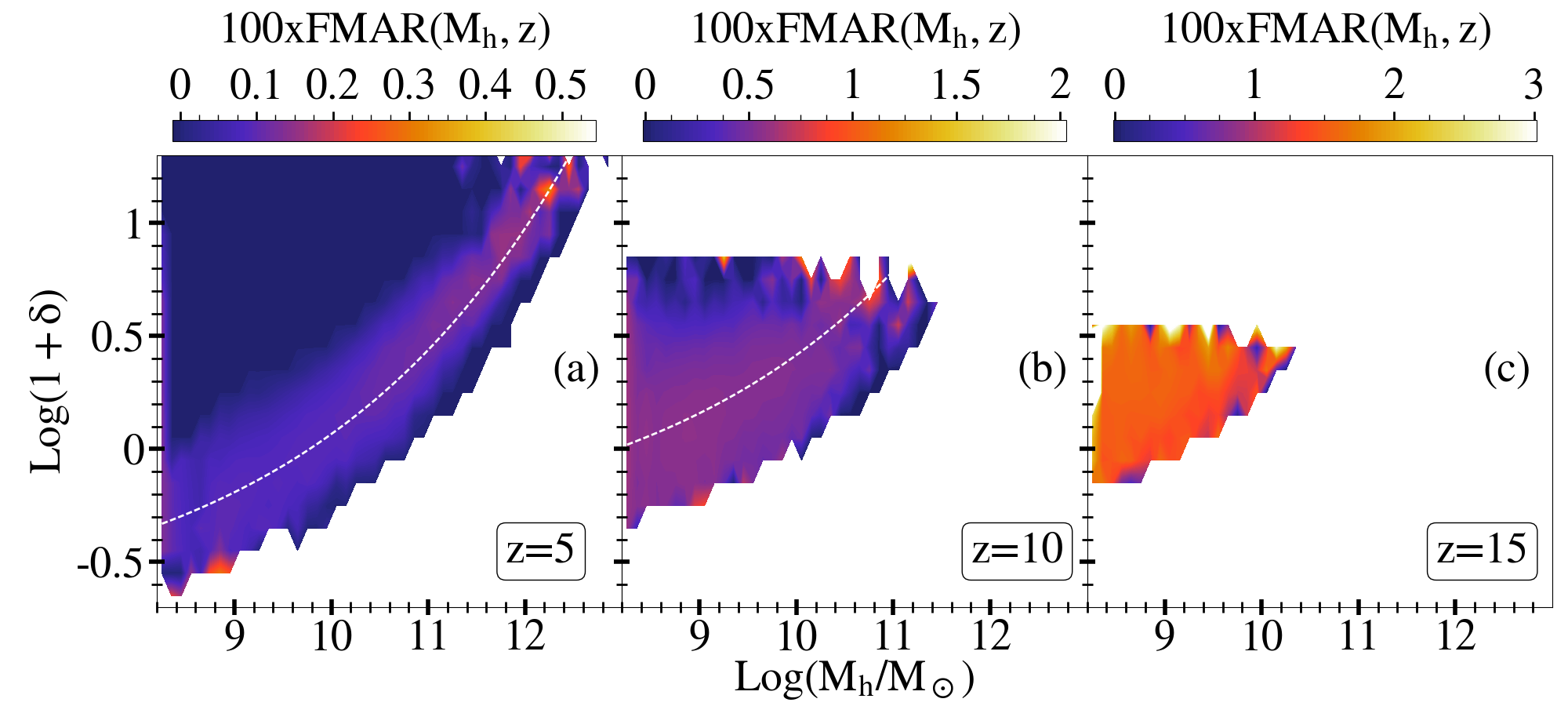}
    \caption{We show the (median) fractional dark matter mass accretion rate (FMAR) as a function of the halo mass and environmental density averaged over $\cube$ at $z=15, 10, 5$, as marked, for the \textit{Photoionization} model. The colorbar shows the FMAR scaled by a 100 for clarity.} 
    \label{fig:dark matter_acc}
\end{figure*}

Focusing on $z=5$, firstly, we see that at fixed halo mass, $R_{\rm vir}/\rhil$ increases with the environmental density smoothed over a cube of $\cube$. For example, for $10^{9.2}\msun$ halos the ratio increases from 0 at $\env \sim -0.5$ to $\sim 2$ for galaxies in environments with $\env > 1$. The same trend persists to higher masses where for $10^{11}\msun$ halos, the ratio increases from 0 at $\env \sim 0.2$ to $\sim 2$ for $\env > 1$. Since the virial radius does not depend strongly on the environmental density, the increase of this ratio is driven by a decrease in the $\rhil$ value with increasing density. This is because, for a given halo mass $m$, it is more probable to find neighboring halos of increasing mass, at roughly the same distance, with an increase in the over-density. To quantify, for $10^{9.2}\msun$ halos with $\env \sim 0$ the most massive neighbors are galaxies with $\mh \sim 10^{11}\msun$ while at $\env \sim 1.5$, the most massive neighbours are up to 15 times more massive, with $\mh \sim 10^{12.2}\msun$. 

Secondly, for low over-densities ($\env \sim -0.5$), we mostly sample field low-mass halos (see also appendix \ref{appendix2}) resulting in a value of $R_{\rm vir}/\rhil \ll 1$. On the other hand, for high-density regions ($\env \sim 1$), the value of $R_{\rm vir}/\rhil$ decreases with increasing halo mass - from about 1.5 for $<10^{10.2}\msun$ halos to about 0 for $10^{12.2} \msun$ halos. This is because while low-mass galaxies in over-dense regions have higher mass neighbours that can influence their accretion, high-mass galaxies in such over-dense regions mostly have lower mass neighbors. 
Moving onto the redshift evolution, we see that at every redshift the halo mass and over-density trends are maintained. i.e. at a given halo mass, the $R_{\rm vir}/\rhil$ value increases with increasing over-density; at a given over-density, the $R_{\rm vir}/\rhil$ value decreases with mass. However, as expected, the range of both the halo masses and over-densities sampled decreases with increasing redshift (c.f. appendix \ref{appendix2}). Further, at a given over-density, we sample increasingly more massive halos with increasing redshift: for example, while $10^{10.2}\msun$ halos in $\env \sim 0.4$ are the most massive halos in their regions at $z=15$ and have $R_{\rm vir}/\rhil \sim 0$, the halos of the same mass at $z=5$ have neighbors with masses up to $10^{11.2} \msun$, which reduces their Hill radius, resulting in $R_{\rm vir}/\rhil \sim 0.5$.

In conclusion, the sphere of influence of a halo is maximum when it is the most massive in its region and progressively decreases with the mass of its neighbors at any redshift. We now examine the consequences of this behavior on the dark matter accretion rates in the next section. 

\subsection{The accretion rates onto dark matter halos} \label{acc_rate}
We now compare the ability of halos to smoothly accrete dark matter from the IGM. To do so, we calculate the fractional mass accretion rate (FMAR) - this is defined as the dark matter mass accreted per Myr expressed as a fraction of the final halo mass such that:

\begin{equation}
    {\rm FMAR}(\mh, z) = \frac{\mhacc(z, z+\Delta z)}{\mh(z) \Delta t} \label{eq:fmar}
\end{equation}
Here, $\mh(z)$ is the final halo mass at $z$, $\mhacc(z, z+\Delta z$) is the halo mass accreted between snapshots $z$ and $z+\Delta z$ and $\Delta t$ is the time difference between these successive snapshots. In Fig. \ref{fig:dark matter_acc} we show the {\rm FMAR} as a function of halo mass and over-density for $z \sim 5-15$.

At $z \sim 5-10$, we see that halos have a mass-dependent "characteristic" environmental density, $\delta_a(z, \mh)$, at which they are the most efficient at smoothly accreting dark matter from the IGM. This is fit by the following relation
\begin{equation}
{\rm log} (1+\delta_a(\mh, z)) = 0.021\times (\mh/\msun)^{0.16} + 0.07 z -1.12.
\end{equation}
This is shown as the dashed (white) line in Fig.~\ref{fig:dark matter_acc}; there is no such clear environmental trend at redshifts as high as $z \sim 15$. As seen, at all $z \sim 5-10$, the characteristic density increases with the halo mass. For example, at $z =5$, ${\rm log} (1+\delta_a)$ increases from about $-0.3$ to $\sim 1.2$ as the halo mass increases from $10^{8.2}$ to $10^{12.2} \msun$. 

At $z=5$ (panel (a) of Fig. \ref{fig:dark matter_acc}), halos in their characteristic environmental density accrete around 0.2\% of their final mass every $\rm Myr$, which at $z=5$ corresponds to accreting $\sim 5\%$ of their mass in the last redshift-step. For such halos, $R_{\rm vir}/\rhil \leq 0.3$ (see e.g.  Fig.~\ref{fig:tidal}), meaning that their sphere of influence is more than three times bigger than their virial radius.

\begin{figure*}
    \centering
    \includegraphics[width=\textwidth]{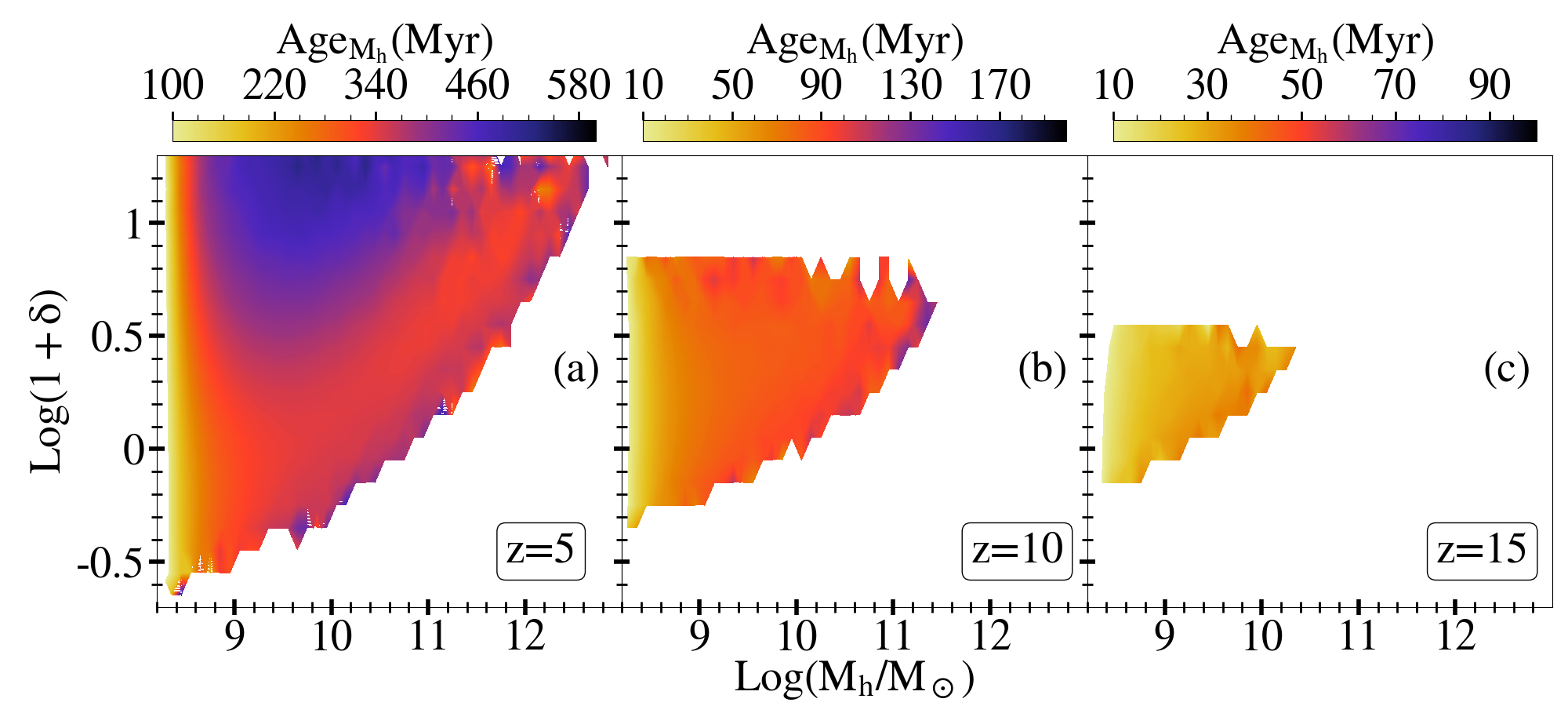}
    \caption{We show the (median) accreted dark matter mass weighted age (${\rm Age_{\mh}}$) as a function of the halo mass and environmental density averaged over $\cube$ at $z=15, 10, 5$, as marked, for the \textit{Photoionization} model. The colorbar shows ${\rm Age_{\mh}}$, as marked.} 
    \label{fig:agemvir}
\end{figure*}

The FMAR then falls off for halos that are either in denser or rarer regions as compared to the characteristic environmental density. This is because two factors play a leading role in determining the accreted dark matter mass: the amount of dark matter in the surrounding IGM available for accretion and its relative velocity with the accreting halo. The amount of dark matter in the surrounding IGM decreases with decreasing density of the environment, which explains that, at fixed halo mass, halos that are in regions less dense than their characteristic environment have a lower fraction of smoothly accreted dark matter mass. Regarding halos in regions denser than their characteristic environment, the increase in dark matter available for accretion is compensated by the increased influence of massive neighbors. Indeed, previous works \citep[e.g.][]{hahn2009} have shown that the velocity shear, created by a massive neighbor, between a halo and the surrounding dark matter in the IGM is an important factor leading to reduced dark matter accretion in dense environments. They also found that it correlates strongly with the tidal field created by the most massive neighbor. Hence, while the amount of dark matter available increases with increasing density of the environment, the combined lower Hill radius and velocity shear created by massive neighbors lead to a stalled accretion for galaxies located in regions denser than the characteristic density.

As we go to $z \sim 10$, firstly, for every halo, the density of the characteristic environment increases by about 0.3 dex. This is the result of both an increased average density of the Universe (which leads to more dark matter available for accretion) and a decrease of $R_{\rm vir}/\rhil$ (indicating that halos can smoothly accrete dark matter from a larger distance). Secondly, the range of environmental density in which halos accrete efficiently increases, from $\sim 0.5$ dex at $z=5$ to the point where there is no clear trend at $z=15$. This is due to the fact that the range of environmental density spanned by halos decreases with increasing redshift. We note that at $z=10$, although the characteristic environment of halos is not as defined as at $z=5$, we can already see the low fraction of smoothly accreted mass of low-mass galaxies in over-dense regions and high-mass galaxies in under-dense regions. Thirdly, galaxies in their characteristic environment accrete more efficiently with increasing redshift as a result of the larger values of the critical density.  

\subsection{The accreted mass-weighted ages in different environments} \label{ages_halos}
We now discuss the dependence of the accreted dark matter weighted ages of halos on their mass and environment as shown in Fig. \ref{fig:agemvir} at $z \sim 5-15$. The mass weighted age at redshift $z$ is calculated as
\begin{equation}
    {\rm Age_{\mh}}(z) = \sum_{z_i=z_0}^z \frac{\mhacc(z_i)}{\mh(z)}(t(z)-t(z_i)) .
\end{equation}
where $\mhacc(z_i)$ is the dark matter mass accreted by all progenitors at redshift $z_i$, $t(z_i)$ is the cosmic time at redshift $z_i$ and $z_0$ is the redshift of formation of the first progenitor.

Starting with $z=5$, from (panel (a) of) Fig.~\ref{fig:agemvir} we see the accreted dark matter weighted age of galaxies in regions with ${\rm log}(1+\delta) \gsim 0.5$ does not depend on halo mass. However, in regions with ${\rm log}(1+\delta) \gsim 0.5$ galaxies that reside in regions corresponding to their characteristic over-density have, on average, lower accreted mass-weighted ages as compared to low-mass galaxies that reside in over-dense regions; while galaxies in characteristic over-density regions have a higher amount of gas available for accretion (leading to a higher FMAR), low-mass galaxies in over-dense regions are inefficient accreters due to the tidal fields from neighboring massive galaxies as discussed in the last sections. For example, halos with $\mh \sim 10^{9.5}$ have a stellar-mass weighted age equal to $\sim 240$ Myr at $\env \sim 0$, while halos of the same mass located in $\env \sim 1.3$ environment have of stellar-mass weighted age almost two times as high, of the order of $\sim 480$ Myr. Finally, we caution against considering the ages for halos at the resolution limit ($\mh \lsim 10^9\msun$) since they only exist in the last few snapshots. 

\begin{figure*}
    \centering
    \includegraphics[width=\textwidth]{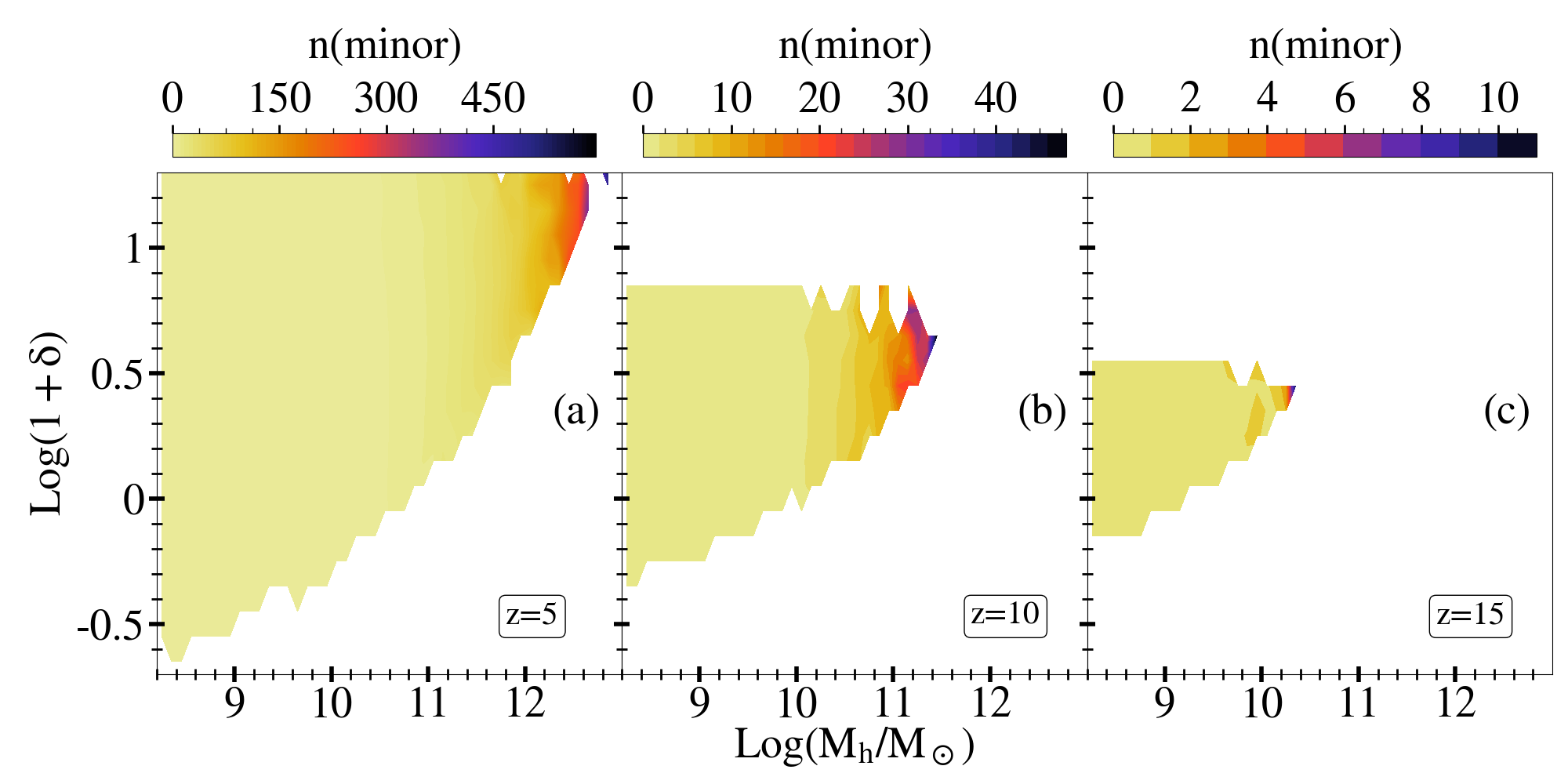}
    \caption{We show the (median) number of minor mergers onto the major branch, ${\rm n (minor)}$, as a function of the halo mass and environmental density averaged over $\cube$ at $z=15, 10, 5$, as marked. The colorbar shows ${\rm n (minor)}$, as marked.} 
    \label{fig:nminor}
\end{figure*}

\begin{figure*}
    \centering
    \includegraphics[width=\textwidth]{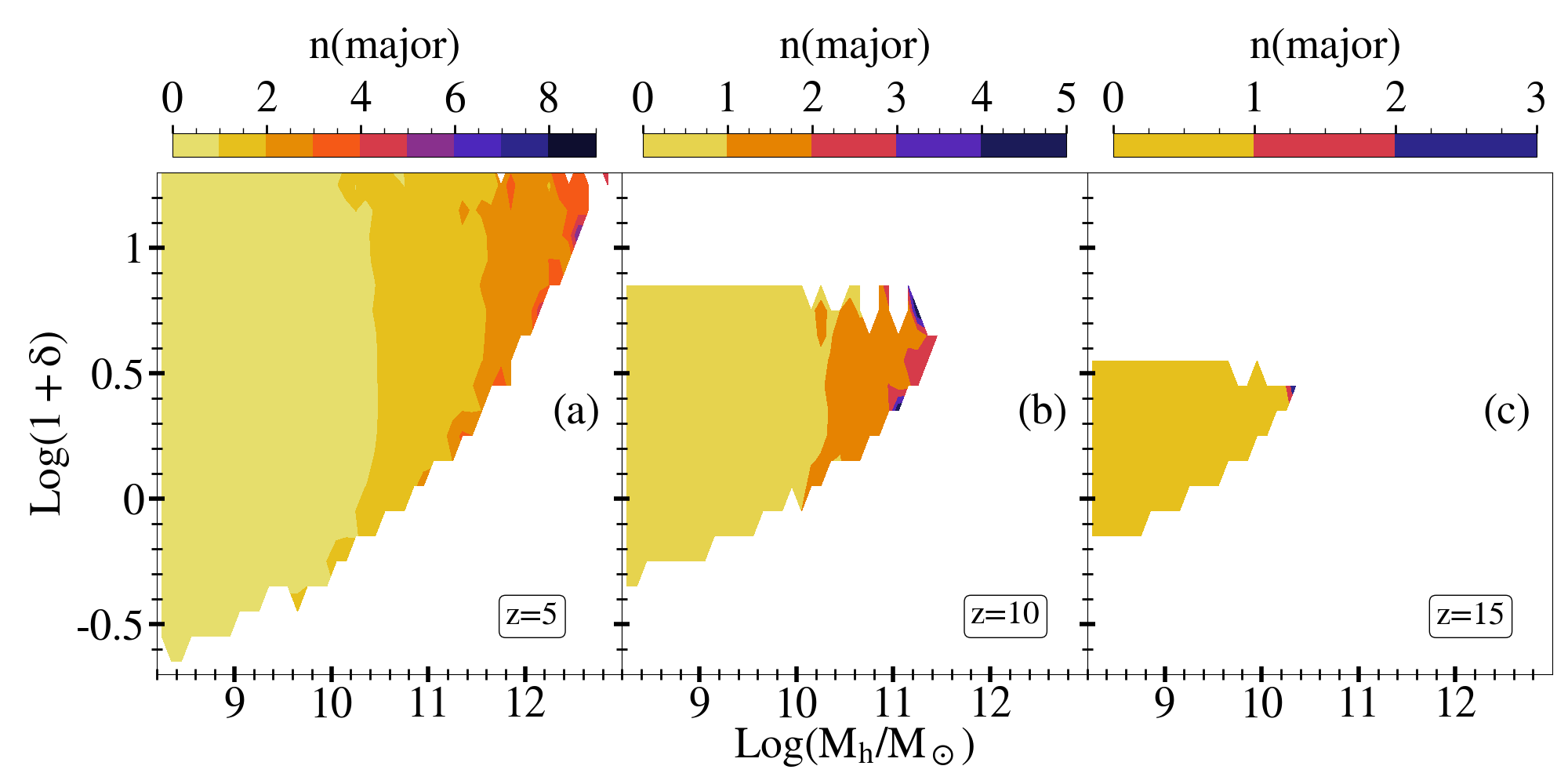}
    \caption{We show the (median) number of major mergers onto the major branch, ${\rm n (major)}$, as a function of the halo mass and environmental density averaged over $\cube$ at $z=15, 10, 5$, as marked. The colorbar shows ${\rm n (major)}$, as marked.} 
    \label{fig:nmajor}
\end{figure*}

\begin{figure*}
    \centering
    \includegraphics[width=\textwidth]{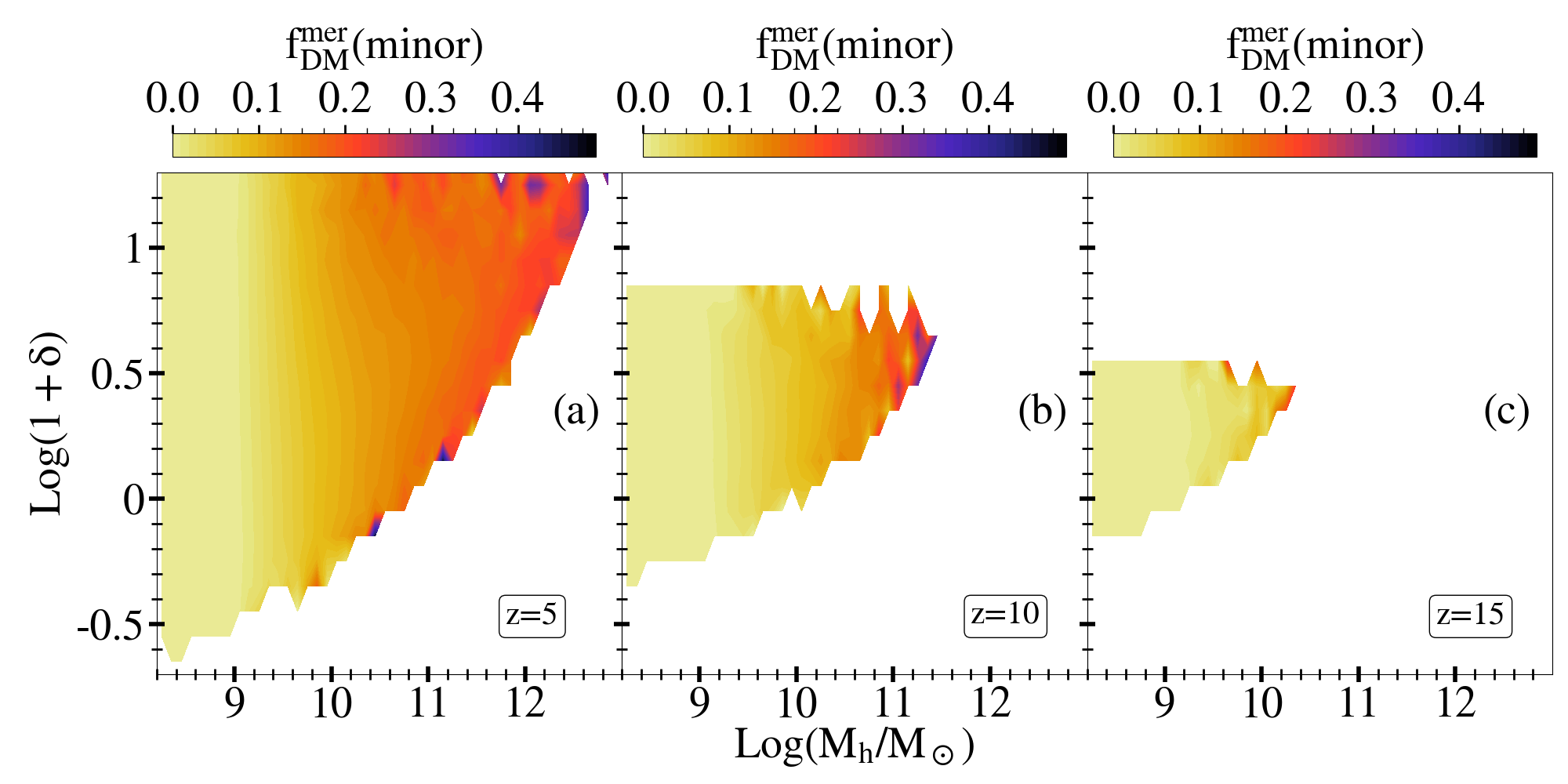}
    \caption{We show the (median) fractional mass brought in by minor mergers onto the major branch, ${\rm f_{DM}^{mer}(minor)}$, as a function of the halo mass and environmental density averaged over $\cube$ at $z=15, 10, 5$, as marked. The colorbar shows ${\rm f_{DM}^{mer}(minor)}$, as marked.} 
    \label{fig:masminor}
\end{figure*}

\begin{figure*}
    \centering
    \includegraphics[width=\textwidth]{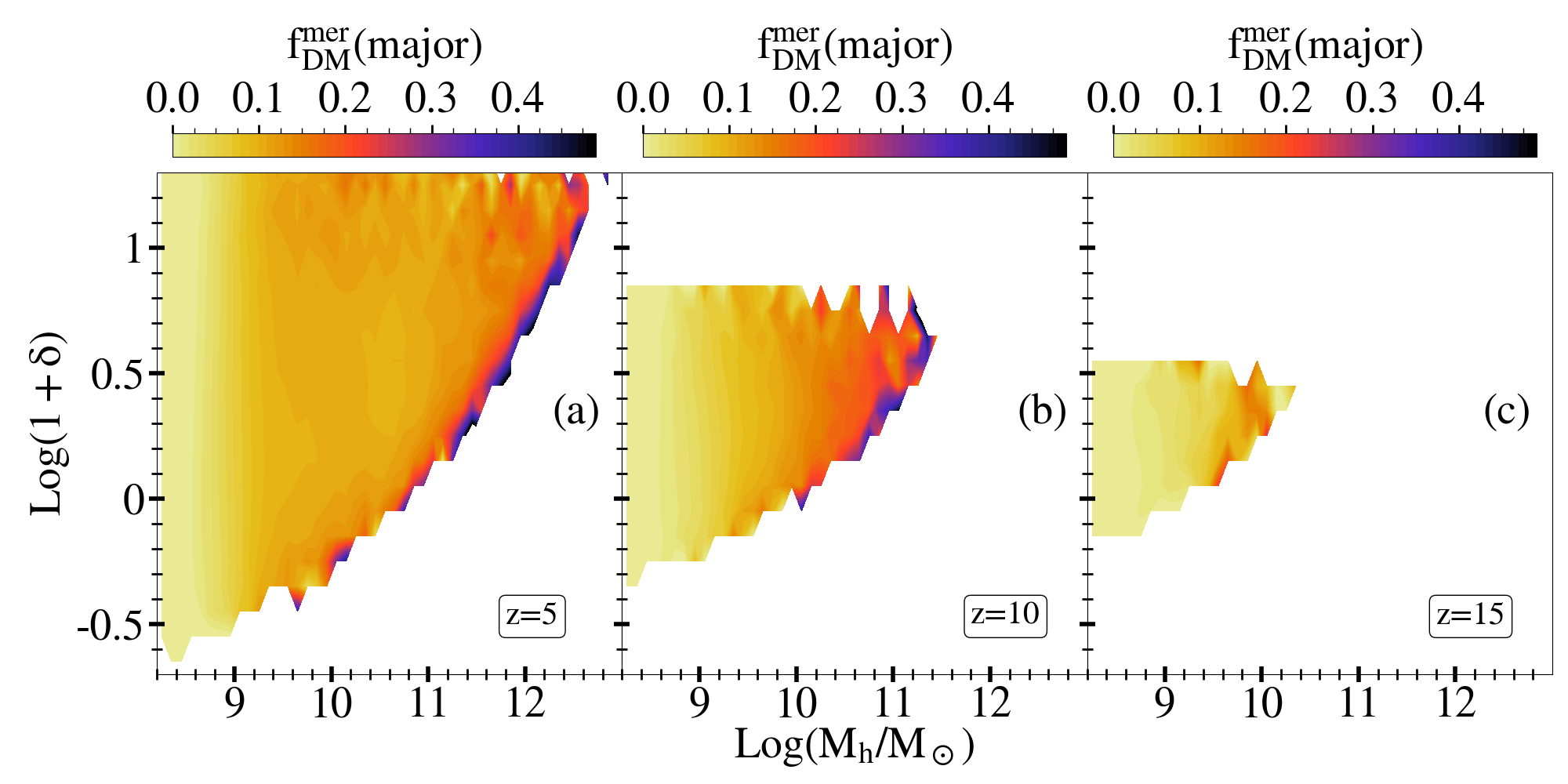}
    \caption{We show the (median) fractional mass brought in by major mergers onto the major branch, ${\rm f_{DM}^{mer}(major)}$, as a function of the halo mass and environmental density averaged over $\cube$ at $z=15, 10, 5$, as marked. The colorbar shows ${\rm f_{DM}^{mer}(major)}$, as marked.} 
    \label{fig:masmajor}
\end{figure*}

\begin{figure*}
    \centering
    \includegraphics[width=\textwidth]{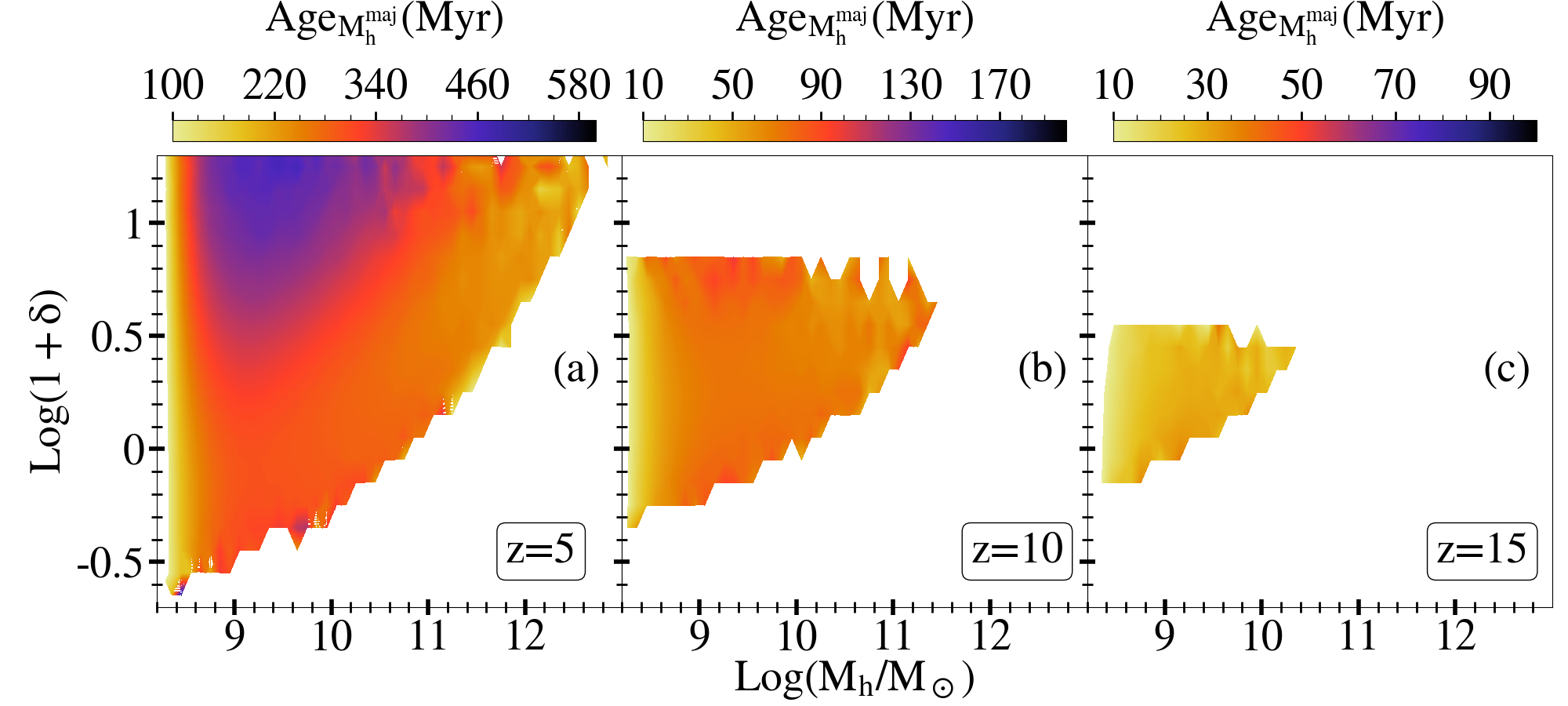}
    \caption{Median dark matter-mass-weighted age of the major branch, ${\rm Age_{\mh}}$, as a function of halo mass and environmental density averaged over $\cube$ at $z=15, 10, 5$ as indicated in the legend in the \textit{Photoionization} model. The colorbar shows ${\rm Age_{\mh}}$, as marked.} 
    \label{fig:fig5}
\end{figure*}

The impact of the environment on the accreted mass-weighted ages becomes weaker as we go to higher redshifts. Although at a given over-density the value of ${\rm Age_{\mh}}$ increases with increasing mass, the trends are much less well defined for galaxies of a given mass at different over-density values. This is possibly driven by the lesser cosmic time available as well as the lower density range sampled by galaxies of a given mass with increasing redshift.

To conclude, we see that, at a given over-density, the age decreases with increasing halo mass, except for $\mh \leq 10^{9}\,\msun$ halos that are at the mass resolution limit. Although this trend seems to be in contradiction with the hierarchical model, it is a result of massive galaxies reducing the accretion of dark matter onto their less massive neighbors in over-dense regions, a trend also observed in simulations, albeit at $z \sim 0$ \citep{hahn2009}.

\subsection{The role of the major branch in halo assembly} \label{major}
In this section, we focus on the assembly of the major branch as a function of the environmental density. We start by discussing the number of minor and major mergers that the major branch undergoes throughout the history of the halo as a function of mass and environment, as shown in Figs. \ref{fig:nminor} and \ref{fig:nmajor}, respectively.
The exact threshold between minor and major mergers is somewhat arbitrary, with e.g. \citet{fakhouri2008} using 1:10, or \citet{genel2010} using 1:3. In the context of studying black hole binaries, \citet{mayer2013} suggest that a mass ratio of 1:5 marks a qualitative difference between minor and major mergers. In this work, we choose to define minor (major) for a halo mass ratio between the secondary and the main progenitors below (above) 1:4, following the definition from \cite{rodriguezgomez2015} applied on halo mass.
As shown in Fig. \ref{fig:nminor}, firstly, we see that, at every redshift, the number of mergers increases with halo mass as expected from the hierarchical assembly. For example, at $z=5$, the number of minor mergers increases from $\sim 1$ at $\mh \sim 10^{9.3} \msun$ to a few hundreds at $\mh \sim 10^{12.5} \msun$. The same trend persists at $z \sim 10$ where the number of minor mergers increases from $\sim 1$ to $40$ as the halo mass increases from $10^{9}$ to $10^{11.5}\msun$. At $z \sim 15$, halos below $\mh = 10^{9.8} \msun$ do not undergo any mergers while halos above that mass experience a few ($\lsim 10$) mergers. Secondly, the number of minor mergers does not show any specific dependence on the environmental density over the entire halo mass range probed for the chosen smoothing volume of $\cube$. 

In Fig. \ref{fig:nmajor}, we see that the number of major mergers too increases with the halo mass. At $z \sim 5$, these increase from $\sim 1$ to $\sim 4$ as the halo mass increases from $\sim 10^{10.5}$ to $10^{12.5}\msun$. As might be expected, the number of major mergers decreases with increasing redshift: indeed, by $z \sim 10$, galaxies with $10^{10.5-11.5}\msun$ experience $\sim 2$ mergers over their entire lifetime while halos at $z \sim 15$ experience one major merger at most. We again find no sensible trend with the underlying environmental density, although we caution that we are limited to the major branch in this calculation. Finally, as expected, the number of minor mergers exceed major mergers by orders of magnitude: $10^{11} \msun$ halos at $z \sim 5 ~ (10)$ undergo $\sim 100 ~ (10)$ minor mergers as compared to $\sim 2-3 ~ (1-2)$ major mergers. 

We then evaluate the amount of mass brought in by both minor and major mergers onto the major branch as a function of both the halo mass and environmental density. As shown in Fig. \ref{fig:masminor}, the importance of the mass brought in by minor mergers increases slightly with halo mass at a given redshift. For example, at $z \sim 5$, minor mergers contribute $\sim 10\% ~ (25\%)$ of the merged mass for the major branch of halos with $\mh \sim 10^{9.5}~ (10^{12})\msun$. The importance of such mergers decreases with redshift such that by $z \sim 10$ they contribute $\lsim 20\%$ of the major branch mass, even for the most massive halos of $10^{11.5}\msun$; there is no discernible mass dependence by $z \sim 15$ with minor mergers bringing in $\lsim 10\%$ of the halo mass. 

As shown in Fig. \ref{fig:masmajor}, the import of major mergers also increases with mass. Despite their lower numbers, they contribute almost the same mass as minor mergers: at $z \sim 5$, such mergers bring in $\sim 10\% ~ (20\%)$ of the major branch mass for $\mh \sim 10^{9.5}~ (10^{12.5})\msun$ halos. Interestingly, major mergers bring in more mass as compared to minor mergers with increasing redshift: at $z \sim 10$, major mergers contribute $\sim 20-40\%$ to the major branch mass for $\mh \gsim 10^{9.5}\msun$ halos (compared to the $\lsim 20\%$ mass fraction of minor mergers), which persists up to redshifts as high as $z \sim 15$ (where minor mergers contribute $\lsim 10\%$ to the major branch mass). 

Unlike the number of mergers, we see at $z\sim 5-10$ that, although the mass brought in by major progenitors is independent of the environment for most halos, halos with $\mh >10^{11.2}\msun$ have a much higher fraction of dark matter mass merged (50\%) in regions less dense than their characteristic environment than in other regions (35\%); this is possibly the result of the major branch having established earlier in low-density regions.

Finally, analogous to the mass weighted age calculated for the entire merger tree in Sec. \ref{ages_halos}, we calculate the mass-weighted age of the major branch as:
\begin{equation}
    {\rm Age}_{\mh^{\rm maj}}(z) = \sum_{z_i=z_0}^z \frac{\mh^{\rm maj,acc}(z_i) + \mh^{\rm maj, mer}(z_i) }{\mh(z)}(t(z)-t(z_i))
\end{equation}
where $\mh^{\rm maj,acc}(z_i)$ and $\mh^{\rm maj,mer}(z_i)$ represent the dark matter mass accreted and merged onto the the major branch at $z_i$. As shown in Fig. \ref{fig:fig5}, we again see the same trends as in Sec. \ref{ages_halos} where the mass-weighted age increases with halo mass for galaxies in regions with $\env \lsim 0.3$ at $z \sim 5$. At higher densities, galaxies that reside in regions corresponding to their characteristic density are younger as compared to galaxies in more over-dense regions: for example, for a halo of $10^{10.5}\msun$, ${\rm Age}_{\mh^{\rm maj}}$ increases from about 340 to 460 Myr as $\env$ increases from 0.5 to 1. Again, the impact of mass weakens as we go to $z \sim 10, 15$. Finally, we find that the mass-weighted age of the major branch is slightly ($\sim 50$ Myr) younger when compared to the assembly of the entire merger tree.

In this section, we have shown that the number of mergers does not depend strongly on the environment and neither does the fraction of mass gained through mergers, with the exception of massive halos located in over-densities below their characteristic environment; these have a much larger fraction of merged mass brought in by their major progenitors.

\section{The dependence of stellar mass assembly on environment and reionization feedback } \label{sec_star}
We now focus on the assembly of stellar mass in different environments for the two limiting radiative feedback scenarios considered in this work. We discuss the role of star formation in both major and minor progenitors in determining the stellar mass in Sec. \ref{stel_assembly} and the impact of environment and reionization feedback on the stellar ages and star formation histories in Secs. \ref{stel_ages} and \ref{stel_sfh}, respectively.

\subsection{The role of star formation in major and minor branches in assembling the stellar content}
\label{stel_assembly}
\begin{figure*}
    \centering
    \includegraphics[width=\textwidth]{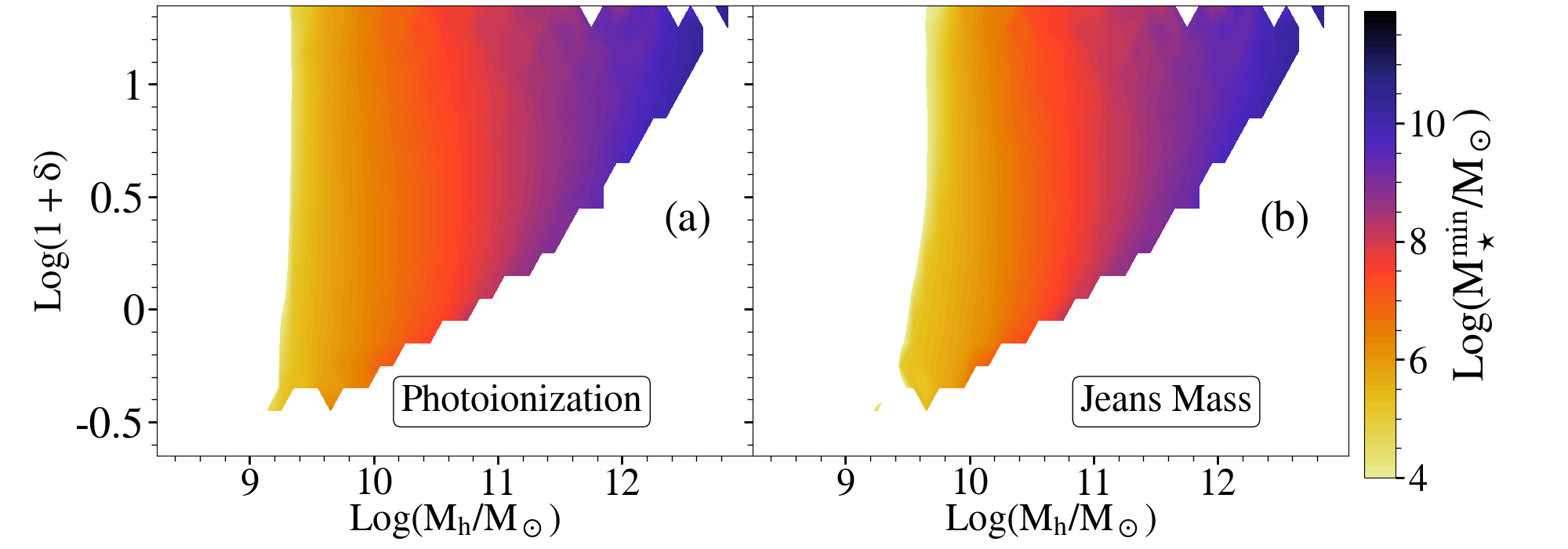}
    \caption{We show the (median) stellar mass formed in minor progenitors that merges onto the major branch over the entire assembly history ($\msmin$) as a function of halo mass and density of the environment averaged over $\cube$ at $z=5$ in the \textit{Photoionization} and \textit{Jeans Mass} models in panels (a) and (b), as marked. The colorbar shows $\msmin$, as marked.} 
    \label{fig:fig10}
\end{figure*}

\begin{figure*}
    \centering
    \includegraphics[width=\textwidth]{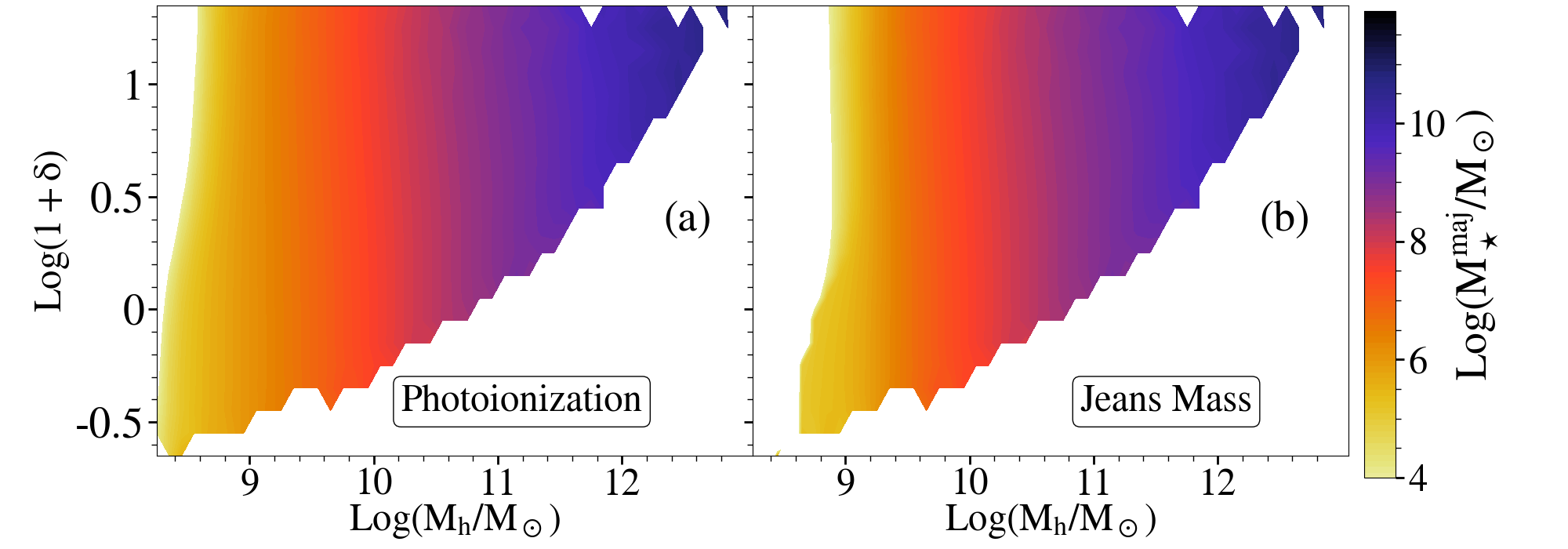}
    \caption{We show the (median) stellar mass formed in major progenitors that merges onto the major branch over the entire assembly history ($\msmaj$) as a function of halo mass and density of the environment averaged over $\cube$ at $z=5$ in the \textit{Photoionization} and \textit{Jeans Mass} models in panels (a) and (b), as marked. The colorbar shows $\msmaj$, as marked.} 
    \label{fig:fig11}
\end{figure*}

We now show the stellar mass assembled across the minor progenitors throughout the assembly history ($\msmin$) of $z \sim 5$ galaxies as a function of halo mass and over-density in Fig. \ref{fig:fig10}. Focusing on panel (a) that shows results for the {\it Photoionization} model, we see that $\msmin$ increases with halo mass, from about $10^5 $ to $10^{10}\msun$ as the halo mass increases from $10^{9.5}$ to $10^{12.5}\msun$. This trend might be expected from the halo mass assembly shown discussed in the sections above. We also see that for this (somewhat minimal) reionization model, $z \sim 5$ halos with masses down to $\sim 10^9\msun$ have star formation in their minor branches. As might be expected from the halo assembly discussed above, we do not find any strong dependence of $\msmin$ with the environmental density for a given halo mass. 

As seen from panel (b) of the same figure, in the \textit{Jeans Mass} model, $z \sim 5$ halos with masses $\lsim 10^{9.4}\msun$ do not have any star formation in their minor branches. This is the result of the strong and instantaneous radiative feedback from reionization in this model that can completely suppress the gas mass of the low-mass progenitors of these halos, quenching star formation. Again, we do not see any strong trend with the environmental density for halos with $\mh \gsim 10^{9.4}\msun$.

We then discuss the assembly of the major branch progenitors in Fig. \ref{fig:fig11}. In the case of the {\it Photoionization} model shown in panel (a), by $z \sim 5$, the progenitors of halos with masses as low as $10^{8.5}\msun$ have formed a stellar mass of $\msmaj \sim 10^{4.5}\msun$ given their larger potentials. Here too, the stellar mass formed increases with the halo mass as might be expected. However, as might be anticipated from the discussion in Sec. \ref{Sec_assembly}, the relative importance of the major and minor branches are mass dependent, although we see no clear indication of an environmental dependence. For example, for low mass halos ($\mh \lsim 10^{9.5}\msun$), the major branch (with $\msmaj \sim 10^8 \msun$) clearly dominates over the minor progenitors ($\msmin \sim 10^5\msun$) in building the stellar content. However, for higher mass halos, both major and minor progenitors contribute roughly equally to the total stellar content: for example, for $\mh \sim 10^{12}\msun$ halos, $\msmin \sim \msmaj \sim 10^{10}\msun$. 

\begin{figure*}
    \centering
    \includegraphics[width=\textwidth]{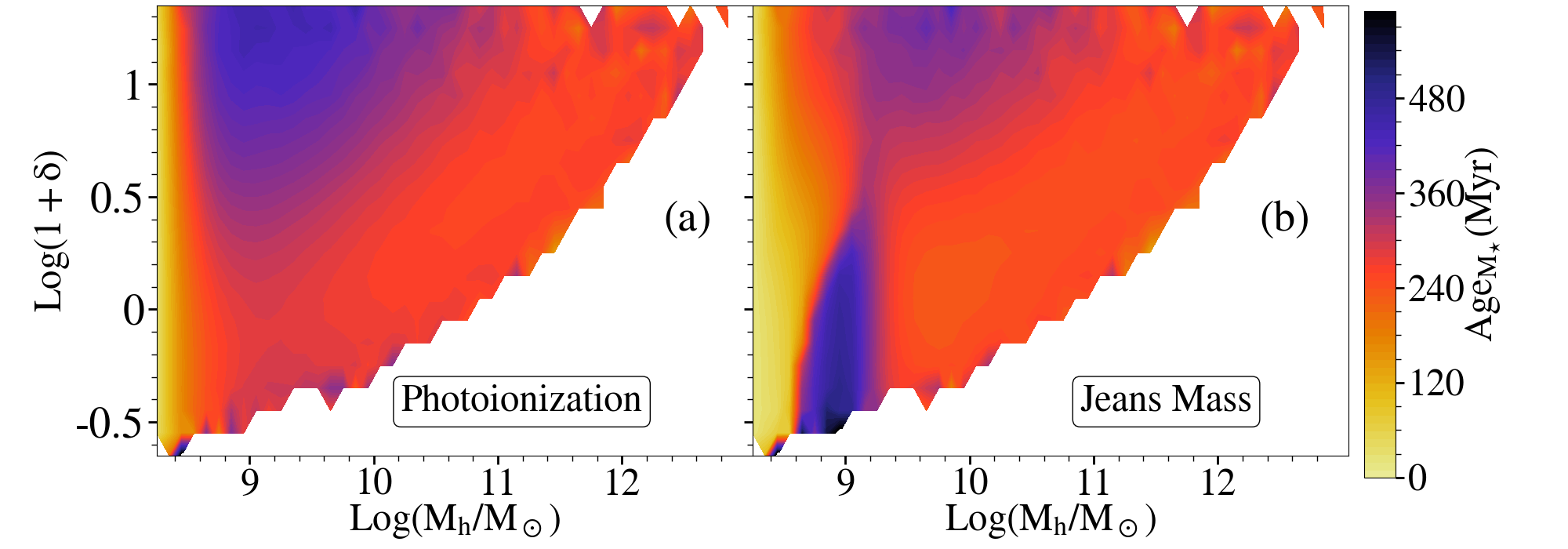}
    \caption{We show the (median) stellar-mass-weighted age of the major branch, $\agems$, as a function of halo mass and environmental density averaged over $\cube$ at $z = 5$ as indicated in the legend in the \textit{Photoionization} model. The colorbar shows $\agems$, as marked.} 
    \label{fig:fig12}
\end{figure*}

Finally, we show the stellar mass assembly in the major branch for the {\it Jeans mass} model in panel (b) of Fig. \ref{fig:fig11}. Given its stronger radiative feedback, in this model, star formation is suppressed in the major branch for increasingly massive galaxies with an increase in the environmental density. For example, halos with $\mh \lsim 10^{8.6} ~ (10^{9})\msun$ are suppressed in terms of star formation for $\env \lsim 0 ~ (\gsim 0.2)$. This is because reionization starts earlier in regions of increasing density which leads to an earlier (instantaneous) suppression of their star formation in this radiative feedback model. As might be expected, the results start converging for $\mh \gsim 10^{9.5}\msun$ halos for both radiative feedback models since their major branch potentials are deep enough so as not to be affected by such feedback.  

\subsection{The impact of environment and reionization feedback on stellar mass-weighted ages}
\label{stel_ages}
We now discuss the stellar mass-weighted ages in Fig.~\ref{fig:fig12}. Starting with the {\it Photoionization} model (panel (a)), the stellar mass-weighted age follows the same trends with the environment as the accreted dark matter -weighted age shown in Fig.~\ref{fig:agemvir}. Namely, $\mh \lsim 10^{10.5}\msun$ galaxies that reside in regions corresponding to their characteristic over-density have, on average, lower stellar mass-weighted ages ($\agems \sim 240$\,Myr) as compared to low-mass galaxies that reside in over-dense regions that are roughly twice as old (with $\agems \sim 480$\,Myr).

The {\it Jeans model} shown in panel (b) of the same figure shows a number of marked differences: below $\mh = 10^{9.4}\msun$, galaxies in low-density environments ($\env \lsim 0.5$) have an older stellar mass in the \textit{Jeans Mass} model than in the \textit{Photoionization} model and this trend reverses as we go to denser environments. In low-density regions, star formation is completely suppressed in the minor branches with the major branch bringing in the older stellar component (Sec. \ref{stel_assembly} above), which pushes up the mass-weighted age. On the other hand, in high-density regions, star formation is suppressed at much earlier times since these get reionized earlier as shown in Fig. \ref{fig:zreion}. This naturally results in younger mass-weighted stellar populations. Hence, up to $10^{11}\msun$, galaxies in low-density environments are younger in the \textit{Jeans Mass} than in the \textit{Photoionization} model. The strong decrease in stellar mass-weighted age for with $\rm M_h \sim 10^{9.2}\,\msun$ in $\env < 0.5$ results from new episodes of star formation as galaxies become massive enough to resist the star formation suppression from radiative feedback.
This stronger suppression of star formation in the \textit{Jeans Mass} model is reminiscent of the prediction of e.g. \citet{barkana2000,barkana2006}, who suggested that a strong reionization feedback would leave an imprint on the star formation histories of low-mass galaxies. Recent cosmological simulations such as the one of \citet{gnedin2014} suggest however that this effect might be fairly weak, closer to the results of our \textit{Photoionization} model.
Above $10^{11}\msun$, there are no differences between the two models.

\begin{figure*}
    \centering
    \includegraphics[width=\textwidth]{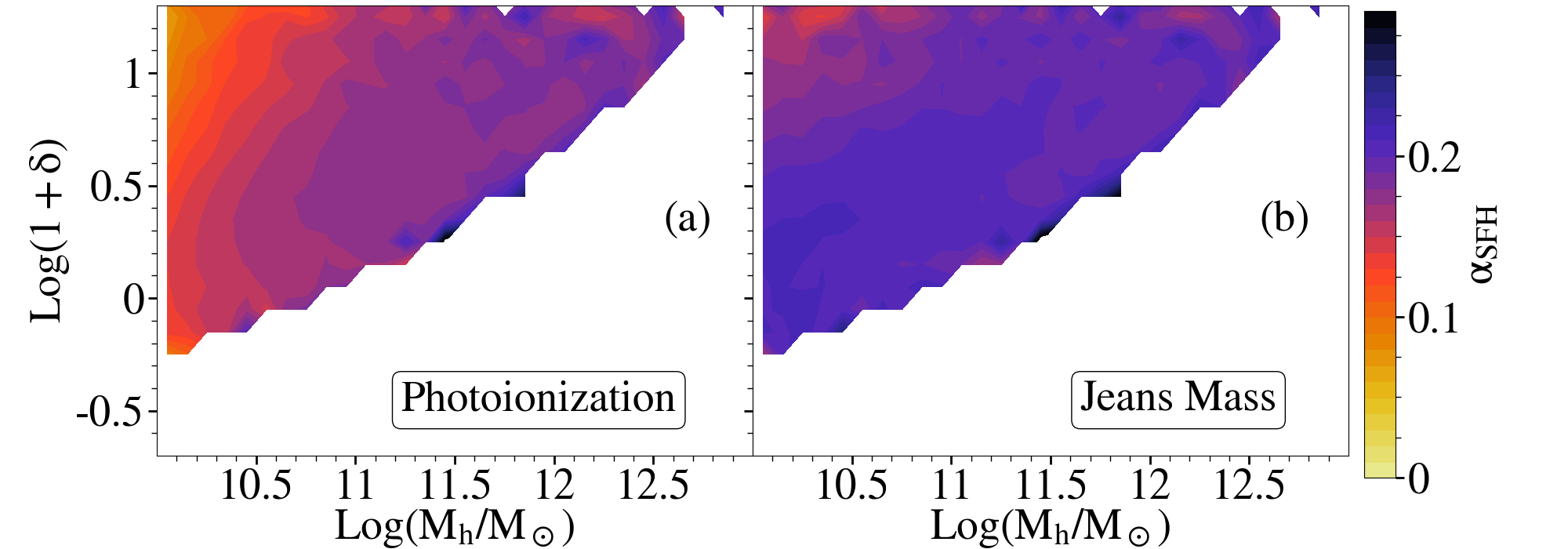}
    \caption{We show the redshift slope of the SFH, $\alpha$, as a function of halo mass and environmental density averaged over $\cube$ at $z=5$ as indicated in the legend in the \textit{Photoionization} model. The colorbar shows $\alpha$, as marked.} 
    \label{fig:sfh}
\end{figure*}

\subsection{The impact of environment and reionization feedback on the star formation history}
\label{stel_sfh}
Finally, we discuss the slope of the star formation history (SFH), as a function of halo mass and density of the environment, in Fig.~\ref{fig:sfh}. Using the same methodology as in \cite{legrand2021}, the SFH for any galaxy is fit as ${\rm Log}({\rm SFR}(z)) = -\alpha (1+z) + \beta$. We note that this fit is valid for stellar masses down to $\sim 10^{8.2}M_\odot$ at $z=5$, which corresponds to a minimum halo mass of $\mh \sim 10^{10}M_\odot$; below this mass, the SFH is too stochastic (i.e. the star formation rate varies rapidly with time) rendering any fit meaningless. 

Starting with the {\it Photoionization} model (panel (a) of Fig. \ref{fig:sfh}), we find that for $\mh \gsim 10^{10.5}\msun$ galaxies, 
$\alpha \sim 0.2$ which is in agreement with our previous results \citep{legrand2021}. In addition, for low-mass galaxies, $\alpha$ decreases with increasing values of $\env$. This is in agreement with the stellar mass-weighted ages discussed in Fig.~\ref{fig:fig12}. Indeed, as galaxies in denser environments have older values of $\agems$ for a given stellar mass, we expect their SFH to become shallower. For a given value of $\env$, the SFH slope steepens with increasing halo mass as these galaxies form stars at a faster rate with time given their increasingly deepening potentials. Finally, we find that $\alpha$ evolves far less (with a value $\sim 0.2$) for most galaxies in the {\it Jeans mass} model as compared to the values in the \textit{Photoionization} model (that vary between 0.1-0.2). This is because the gas mass (and hence the star formation rate) for galaxies is affected over a much longer timescale in the \textit{Photoionization} model as compared to the instantaneous feedback in the {\it Jeans mass} model, leading to a steeper slope for the latter. This behavior also explains part of the scatter in the SFH slope shown in Fig. 8 of \cite{legrand2021}.

\section{Conclusions \& discussion} \label{Sec_conclusion}

In this work, we study the dependence of the assembly of galaxies and their baryonic component on the density of their environment and the strength of radiative feedback during the Epoch of Reionization. For this purpose, we use the \textsc{astraeus} framework, a combination of N-body simulation, semi-analytic galaxy evolution model and self-consistent semi-numerical radiative feedback scheme which has a box size of $160\,h^{-1}$Mpc, a particle mass resolution of $6.2\times10^6\, h^{-1}\rm M_\odot$ and reproduces the key observables for galaxies in the Epoch of Reionization. To assess the role of the reionization, we consider two radiative feedback scenarios: the \textit{Photoionization} model is a time-delayed, weak radiative feedback while the \textit{Jeans Mass} model is an instantaneous maximum radiative feedback. Our main results are:

\begin{enumerate}
    \item The influence of a halo over the surrounding DM and gas in the IGM is at its strongest when it is the most massive halos in its neighborhood and its influence decreases if the halo has massive neighbors due to the tidal field of the latter. As the number of more massive neighbors is linked to the environmental density, this results in a characteristic density of the environment at which they are the most efficient at accreting dark matter, which follows ${\rm log} (1+\delta_a(\mh, z)) = 0.021\times \mh^{0.16} + 0.07 z -1.12$, up to $z\sim 10$. At higher redshifts, the small range of over-density spanned by halos makes the trend unclear. 
    \item Due to their increased dark matter accretion, halos at $z=5$ located in their characteristic environment have an accreted dark matter age up to twice as low ($\sim 240\,$Myr) as halos of similar mass in denser environments ($\sim 480\,$Myr). The dependence of the accreted dark matter age on the density of the environment weakens as we go to higher redshifts (Fig.~\ref{fig:agemvir}).
    \item The number of minor (<1:4) or major (>1:4) mergers undergone by the major branch throughout the history of a halo increases with increasing halo mass, e.g at $z=5$ from $\sim 1$ minor merger and no major merger for $10^{9.3}\msun$ halos to hundreds of minor mergers and $\sim 4$ major mergers for $\sim 10^{12}\msun$ halos and is independent of the environment (Fig.~\ref{fig:nminor} and \ref{fig:nmajor}). The fraction of dark matter brought in by minor mergers is slightly higher (up to 10\%) then the one brought in major mergers for most halos with $\mh \gtrsim 10^{9.5}\msun$ at $z \sim 5$, with the two quantities being similar at higher redshift. In addition, the DM mass brought in by either type of merger is mostly independent of the environment except for massive galaxies in low-density environments, for which major mergers bring much more mass, of the order of 50\% (Fig.~\ref{fig:masmajor}). 
    \item Following the same trends, the fraction of stellar mass brought in by both minor and major mergers increases with halo mass. However the contribution from major mergers dominates only for low-mass galaxies ($\mh \lesssim 10^{9.5}\msun$) and equates the contribution from minor mergers at higher masses ($\mh \sim 10^{12}\msun)$ (Fig.~\ref{fig:fig10} and  \ref{fig:fig11}). The effect of radiative feedback on the mass assembly is restricted to galaxies with $\mh \lesssim 10^{9.5}\msun$ for which a stronger radiative feedback can suppress the contribution from minor mergers (Fig.~\ref{fig:fig12}).
    \item The radiative feedback also affects the stellar mass-weighted age: while in both models the stellar mass-weighted age follows the same trends as the accreted dark matter mass-weighted age (at a given mass, galaxies in dense regions are older than their counterparts in low-density environments), galaxies of up to $10^{11}\msun$ in the \textit{Jeans Mass} have lower stellar mass-weighted age than in the \textit{Photoionization} model due to the earlier suppression of star formation in the \textit{Jeans Mass}.
    \item The slope of the star formation history is also affected by the radiative feedback: due to the dependence on the time of reionization (and hence on the environment) of the radiative feedback strength in the \textit{Photoionization} model, the SFH of galaxies with $10^{10}\msun < \mh < 10^{11.5}\msun$ becomes shallower with increasing density of the environment, its slope decreases from $\sim0.18$ in $\env \sim 0.3$ environments to $\sim 0.1$ in $\env \sim 1.2$ environments  (Fig.~\ref{fig:sfh}).
\end{enumerate}

It is important to note that the work presented has a few caveats. First, our results depend on the scale over which the density field is smoothed out. In this work, we smooth out the density on a scale of $\cube$. Smoothing the field over too big a scale leads to homogenizing the field to the point that environmental trends are non-existent while reducing it leads to a strong bias between the density of a cell and the most massive halo within that cell. Other measures of the environment exist such as the in \cite{fakhouri2009}, where they subtract out the FOF mass of the central halo within a sphere.
Second, for galaxies with $\mh < 10^{10.2}\msun$, we do not resolve mergers with ratios down to 1:100. It would be beneficial to extend the study to lower-mass galaxies in order to confirm the result for galaxies with $\mh < 10^{10.2}\msun$, going down to $10^7\msun$ would allow accounting for mergers up to a ratio of 1:100, possibly increasing the mass brought by mergers. 
Third, the evolution of galaxies in our model is tied to the halo potential so galaxy properties such as the star formation rate are strongly driven by the gravitational potential. 
Lastly, we assume a constant escape fraction of ionizing photons in both models explored. Assuming a mass-dependent escape fraction would lead to a different reionization topology and therefore a different radiative feedback, especially on low-mass halos.  

The results of this work could in principle be tested observationally: indeed, as we find that the environment and the strength of the reionization feedback affect the star formation histories of low-mass galaxies, both effects should leave an observable imprint. For instance, deep JWST observations will probe the star formation histories of galaxies in different environments (e.g. through star formation rate measurements and stellar ages estimations). To disentangle also between the effect of environment and radiative feedback, however, these will need to be correlated with probes of the (local) reionization history, such as the transmission of the Lyman-$\alpha$ line or 21 cm observations with SKA. We will address the connection between the reionization history and the Lyman-$\alpha$ transmission in the next paper of this series \citep{hutter2022}.
The number of observed high-redshift galaxies is expected to rise significantly with the launch of Next-generation facilities such as the James Webb Space Telescope or the Nancy Grace Roman Space Telescope. The work presented in this paper will be useful to understand the assembly of these galaxies and the resulting properties.

\section*{Acknowledgements}
We wish to thank the anonymous referee for comments that improved this paper. We are grateful to Christopher Lovell for valuable comment. LL, PD, AH, SG, MT and GY acknowledge support from the European Research Council's starting grant ERC StG-717001 (``DELPHI"). PD, AH, SG, MT and GY acknowledge support from the NWO grant 016.VIDI.189.162 (``ODIN") and PD acknowledges support from the European Commission's and University of Groningen's CO-FUND Rosalind Franklin program.
GY acknowledges financial support from MINECO/FEDER under project grant AYA2015-63810-P and MICIU/FEDER under project grant  PGC2018-094975-C21.
The authors wish to thank V. Springel for allowing us to use the L-Gadget2 code to run the different Multidark simulation boxes, including the VSMDPL used in this work. The VSMDPL simulation has been performed at LRZ Munich within the project pr87yi. The CosmoSim database (\url{www.cosmosim.org}) provides access to the simulation and the Rockstar data. The database is a service by the Leibniz Institute for Astrophysics Potsdam (AIP).

\section*{Data Availability}
The \textsc{Astraeus} simulations and derived data in this research will be shared on reasonable request to the corresponding authors.


\bibliographystyle{mnras}
\bibliography{mybib}



\appendix
\section{Dependence of the reionization redshift on halo mass and density of the environment}
\label{appendix1}

\begin{figure*}
    \centering
    \includegraphics[width=\textwidth]{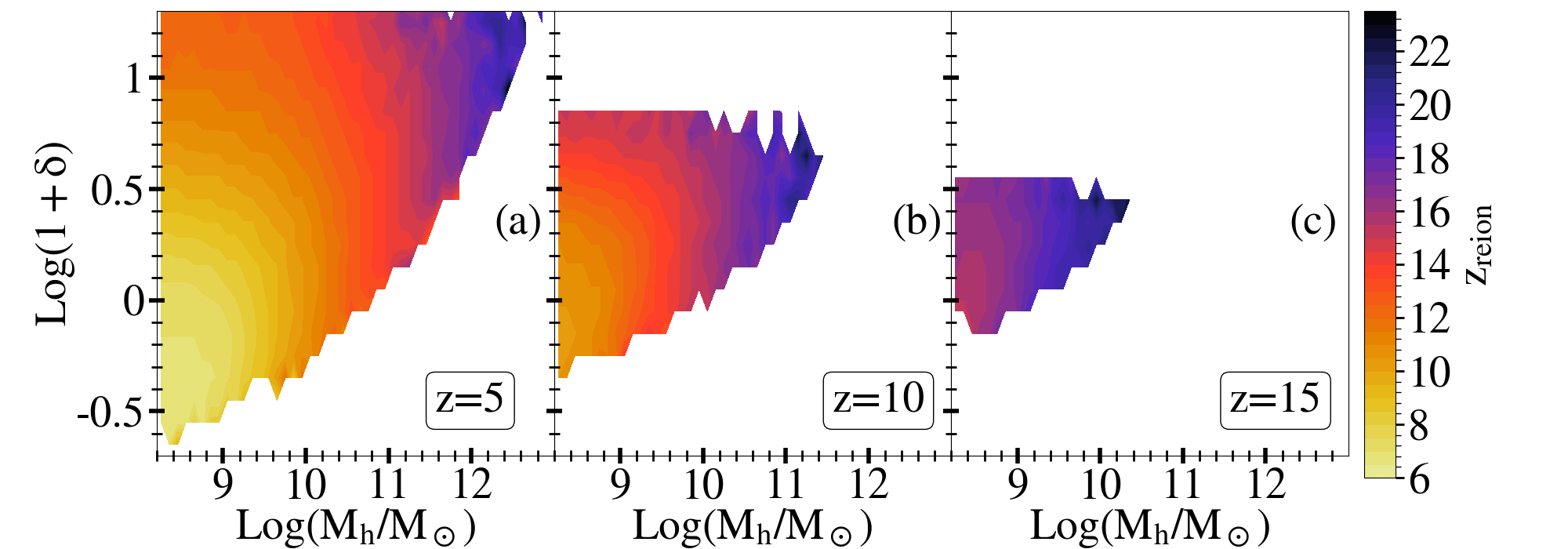}
    \caption{Median reionization redshift (${\rm z_{reion}}$) as a function of halo mass and density of the environment averaged over $\cube$ at $z=15, 10, 5$, as indicated, the \textit{Photoionization} model. The colorbar shows ${\rm z_{reion}}$ as marked.}
    \label{fig:zreion}
\end{figure*}

In Fig.~\ref{fig:zreion}, we show the reionization redshift ($z_{\rm reion}$) of galaxies as a function of their halo mass and the environmental density at $z \sim 5-15$ in the \textit{Photoionization} model. 

At a given value of $\env$, $z_{\rm reion}$ increases with increasing halo mass. This is partly due to the averaging of the environment density over $cube$: although they have the same resulting average environment density, low-mass halos are in on the outskirt of dense regions while high-mass galaxies are in the center. In the inside-out reionization scenario, the high-mass galaxies will ionize their cells first before the ionizing photon can escape and start ionizing surrounding cells with similar environment density but containing mostly low-mass galaxies. At a fixed halo mass, as we move to denser regions the ionization scenario change to an inside-out topology, in which regions get reionized by the galaxies they contain. Thus, we see that the reionization redshift increases with increasing density: while most cells with $\env < 0$ gets reionized  between $z=7$ and $z=6$, cells with $\env > 1$ gets reionized as early as $\sim 12$ and some of the densest region get reionized at $z \sim 20$. Although a denser region means a higher amount of neutral hydrogen and recombination that would lead to a later reionization, the presence of very massive galaxies that are efficient at ionizing their surroundings compensate and denser regions end up being reionized earlier. Finally, although the radiative feedback is different between the two radiative feedback models studied here, the reionization redshift does not strongly depend on the model adopted, i.e the topology of reionization is mostly independent of the radiative feedback scheme.

\section{Distribution of dark matter halos as a function of mass and environment}
\label{appendix2}

\begin{figure*}
    \centering
    \includegraphics[width=\textwidth]{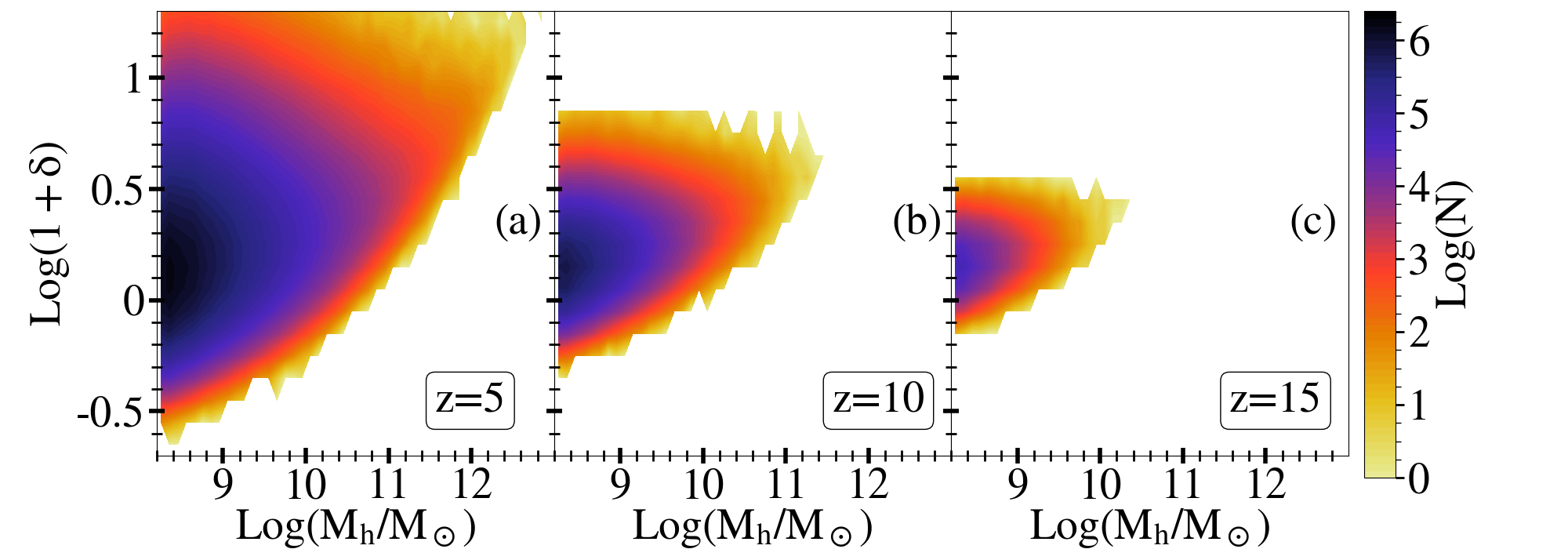}
    \caption{Number of galaxies (N) as a function of halo mass and density of the environment averaged over $\cube$ at $z=15, 10, 5$, as indicated. The colorbar shows the value of N as marked.} 
    \label{fig:logN}
\end{figure*}

In Fig.~\ref{fig:logN}, we show the number of halos in each halo mass bin as a function of the environmental density at $z \sim 5-15$. While the distribution of halos follow the same trend at all redshifts, both the range of halo masses and over-densities probed increases with decreasing $z$: while halos at $z=15$ have halo mass of at most $\mh \sim 10^{10.2} \msun$ and are located in environments with density $-0.2 < \env < 0.5$, the mass of halos at $z=5$ extends to $\mh \sim 10^{12.8}\msun$ and are located in environments with  $-0.6 < \env < 1.3$. 
At each redshift, the bulk of halos have low-masses ($\mh \lsim 10^9 \msun$) and are located in medium-density environments ($-0.2 < \env < 0.5$). As we go to higher halo mass, the number of halos decreases, in agreement with the hierarchical model, such that halos of increasing mass can only be found in increasingly denser environments, e.g while halos with $\mh \sim 10^9\msun$ can be found in almost all environments ($\env > -0.5$), massive halos with $\mh > 10^{12}\msun$ are limited to regions with $\env > 0.5$. Further, at a fixed halo mass, the number of halos falls off for both highly under- and over-dense regions. This is due to a combination of the paucity of low-mass halos in under-dense regions and over-dense regions mostly hosting massive halos.


\bsp	
\label{lastpage}
\end{document}